\documentclass[aps,pra,twocolumn,amsmath,amssymb,nofootinbib,superscriptaddress]{revtex4}

\newcommand{\bra}[1]{\langle#1|}
\newcommand{\ket}[1]{|#1\rangle}

\usepackage{graphicx}
\usepackage{mathrsfs}
\usepackage[colorlinks]{hyperref}

\newcommand{\Fcond}{\tilde{\mathcal{F}}}
\newcommand{\Funco}{\bar{\mathcal{F}}}

\begin{document}

\bibliographystyle{apsrev}

\title{Multiplexed single-photon state preparation using a fibre-loop architecture}

\author{Peter P. Rohde}
\email[]{dr.rohde@gmail.com}
\homepage{http://www.peterrohde.org}
\affiliation{Centre for Engineered Quantum Systems, Department of Physics \& Astronomy, Macquarie University, Sydney NSW 2113, Australia}

\affiliation{Centre for Quantum Computation and Intelligent Systems (QCIS), Faculty of Engineering \& Information Technology, University of Technology, Sydney, NSW 2007, Australia}

\author{L. G. Helt}
\affiliation{Centre for Ultrahigh bandwidth Devices for Optical Systems (CUDOS), MQ Photonics Research Centre, Department of Physics \& Astronomy, Macquarie University, NSW 2109, Australia}

\author{M. J. Steel}
\affiliation{Centre for Ultrahigh bandwidth Devices for Optical Systems (CUDOS), MQ Photonics Research Centre, Department of Physics \& Astronomy, Macquarie University, NSW 2109, Australia}

\author{Alexei Gilchrist}
\affiliation{Centre for Engineered Quantum Systems, Department of Physics \& Astronomy, Macquarie University, Sydney NSW 2113, Australia}

\date{\today}

\frenchspacing

\begin{abstract}
Heralded spontaneous parametric down-conversion (SPDC) has become the mainstay for single-photon state preparation in present-day photonics experiments. Because they are heralded, in principle one knows when a single photon has been prepared. However, the heralding efficiencies in experimentally realistic SPDC sources are typically very low. To overcome this, multiplexing techniques have been proposed which employ a bank of SPDC sources in parallel, and route successfully heralded photons to the output, thereby effectively boosting the heralding efficiency. However, running a large bank of independent SPDC sources is costly and requires complex switching. We analyse a multiplexing technique based on time-bin encoding that allows the heralding efficiency of just a single SPDC source to be increased. The scheme is simple and experimentally viable using present-day technology. We analyse the operation of the scheme in terms of experimentally realistic considerations, such as losses, detector inefficiency, and pump-power.
\end{abstract}

\maketitle

\section{Introduction}

Single-photon state preparation has numerous applications in the field of quantum photonics. Most notably, it is an essential requirement for optical quantum information processing protocols, such as linear optics quantum computing \cite{bib:KLM01}, boson-sampling \cite{bib:AaronsonArkhipov10}, and optical quantum metrology \cite{bib:MORDOR}, which require high-efficiency heralded single-photon sources.

Spontaneous parametric down-conversion (SPDC) has become the \emph{de facto} standard for generating heralded single-photon states, owing to the relative simplicity of its experimental implemention. However, single source SPDC suffers from being non-deterministic, and generating a sequence of photons in this manner is exponentially inefficient. To overcome this problem previous authors have proposed and studied various schemes for combining multiple independent SPDC sources, making use of both spatial \cite{14adam053834,bib:MosleyMultiplex,bib:ObrienMultiplex,14meany42,bib:CollinsMultiplex,13mazzarella023848,bib:ChristMultiplex,bib:ZeilingerMultiplex,bib:JenneweinBarbieriWhite,bib:ShapiroMultiplex,bib:MigdallMultiplex} and temporal \cite{15mendoza,14adam053834,13schmiegelow447,13glebov031115,11mower052326,08mccuskerjtua117,06peters630507, 04jeffrey1, 04pittman57,bib:PittmanSPDC,bib:Nunn13} multiplexing.

Recently a fibre-loop architecture for implementing boson-sampling was presented \cite{bib:MotesLoop}, where the input state is time-bin-encoded, and multi-photon interference is implemented via dynamically-controlled beamsplitters. Subsequently it was shown \cite{bib:RohdeLoopLOQC} that the same architecture can be modified to implement universal linear optics quantum computing. 

The viability of loop and delay-line based architectures has previously been experimentally demonstrated in the context of multiplexed, number-resolved photo-detection \cite{bib:Fitch03, bib:Achilles04, bib:RohdeWebb07}, quantum random walks \cite{bib:Schreiber10, bib:Schreiber12}, and quantum memory \cite{bib:Pittman02}. These demonstrations show that temporally-encoded schemes are capable of maintaining quantum coherence over realistic delay lengths with present-day technology.

Here we show that a simplified loop-based architecture can implement multiplexed single-photon state preparation, using time-bin encoding. The scheme requires only a single SPDC source with high repetition rate, a single fibre-loop, a dynamically controlled switch, and a single photo-detector with time-resolution on the order of the time-bin separation. The SPDC source is operated at a high repetition rate to minimise the length of the fibre-loop, and thus losses. This protocol allows a source to be constructed with very high heralding efficiency and fidelity, limited only by loss rates in the switch and fibre-loop, and the detector efficiency. The experimental requirements to build this architecture are largely available today, and some of the key elements have previously been experimentally demonstrated.

Using a single instance of the source, loop, and detector immediately confers some advantages over a multiplexed array of such elements as we do not have to manufacture each element identically to ensure that the photons generated are indistinguishable. Generating indistinguishable photons is a stringent requirement for a source suitable for quantum information processing.

The concept of the scheme is to employ a single SPDC source, repeatedly triggered at a high repetition rate, yielding a correlated pulse-train in  two output spatial modes. The pulse-train is temporally multiplexed, such that a successfully heralded photon is routed to one of the output time-bins, which then closely approximates a single-photon source.

We build on a circuit layout employed by Pittman, Jacobs \& Franson \cite{bib:PittmanSPDC}, extending  their results by explicitly deriving the heralding efficiency and fidelity of the prepared state, taking detector efficiency, and fibre and switch loss rates into account. Additionally, we describe the optimal switching sequence to employ, so as to maximise fidelity. We further consider the steady state behaviour of the device, which allows single-photon preparation to take place with asymptotically close to unit heralding efficiency, and allows immediate extraction of a stored photon, without having to wait for a multiplexing sequence to complete (i.e. an on-demand `push-button' source).

\section{Architecture}

\begin{figure}[!htb]
\includegraphics[width=0.7\columnwidth]{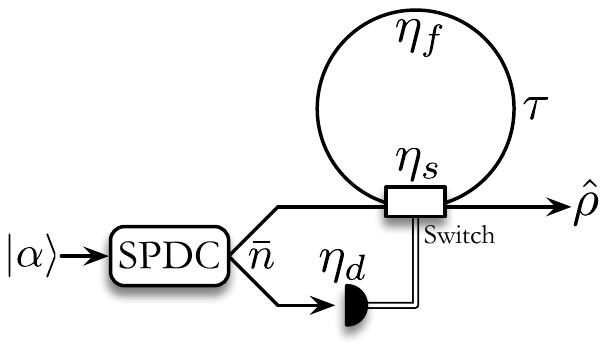}
\caption{Architecture for multiplexed single-photon state preparation, as first described by Pittman, Jacobs \& Franson \cite{bib:PittmanSPDC}. A single SPDC source is pumped with a coherent state, $\ket\alpha$, with high repetition rate, preparing a pulse-train with time-bin separation $\tau$. One port of the SPDC source is monitored with a photo-detector (either number-resolving or `bucket', i.e. on/off), which must have time-resolution better than $\tau$ so as to distinguish the time-bins. The other port is coupled to a fibre-loop of length $\tau$, with a dynamically-controlled on/off switch that can be switched within time $\tau$. The control sequence of the switch is chosen according to the measurement signatures in the detected mode. Here $\eta_s$, $\eta_f$ and $\eta_d$ are the efficiencies of the switch, fibre-loop, and detector respectively. At the output we have some state $\hat\rho$, which describes a pulse-train in which the last time-bin ought to closely approximate a single-photon Fock state.} \label{fig:arch} 
\end{figure}

The full architecture for our protocol is shown in Fig.~\ref{fig:arch}. The goal is to prepare an output pulse-train in which the last time-bin in a sequence closely approximates the single-photon Fock state, $\ket{1}$. The pulse-train in one mode is measured using time-resolved photo-detection, whilst the other enters a loop architecture comprising a single fibre-loop, and a dynamically controlled switch. The dynamic switch need only toggle between completely reflective and completely transmissive. The switching sequence is controlled by the measurement signatures in the measured mode. Specifically, the switch settings at each time-bin are chosen so as to route the most-recent successful SPDC heralding event to the last output time-bin. Then, in the ideal case, that time-bin will contain a single-photon if at least one SPDC heralding event was successful.


Our scheme implements a type of multiplexed state preparation using far more frugal resources compared to spatially multiplexed schemes --- requiring only a single SPDC source, a single dynamically controlled switch, and a single time-resolved photo-detector. The architecture presents a significant efficiency improvement over stand-alone SPDC sources without multiplexing, and the components are largely available today.

Below, we explicitly take into account the effects of the different losses in the architecture (loss being the dominant experimental obstacle), as well as considering both number-resolved and `bucket' (or on/off) detectors. The latter is standard in present-day experiments, whilst the former is presently very challenging and expensive, but is becoming more common. The theoretical treatment we present could easily be extended to incorporate other error models, such as dark-counts.

\section{Photo-detection}

In the photon-number basis, a number-resolved photo-detector may be characterised in terms of a matrix of conditional probabilities --- the probability of measuring $m$ photons, given that $n$ photons were incident on the detector. For a number-resolved detector with efficiency $\eta_d$, these conditional probabilities are given by
\begin{eqnarray}
p_\mathrm{det}(m|n) = \binom{n}{m} {\eta_d}^m (1 - \eta_d)^{n-m}.
\end{eqnarray}
The $p_\mathrm{det}(m|n)$ matrix completely characterises the operation of the detector in the photon-number basis, and can easily be modified to include other effects, such as dark-counts, which we do not treat here.

In the architecture we present, only two measurement outcomes are of interest --- a single photon is detected (\mbox{$m=1$}), or no photons are detected (\mbox{$m=0$}). In this instance the conditional probabilities are given by
\begin{eqnarray}
p_\mathrm{det}(0|n) &=& (1 - \eta_d)^n, \nonumber \\
p_\mathrm{det}(1|n) &=& \eta_d (1 - \eta_d)^{n-1} n.
\end{eqnarray}

For bucket detectors, which can only distinguish between zero photons and more than zero photons, the conditional probabilities of the two measurement outcomes are given by
\begin{eqnarray}
p_\mathrm{det}(\mathrm{no\,click}|n) &=& p_\mathrm{det}(0|n) \nonumber \\
&=& (1 - \eta_d)^n, \nonumber \\
p_\mathrm{det}(\mathrm{click}|n) &=& 1 - p_\mathrm{det}(0|n) \nonumber \\
&=& 1 - (1 - \eta_d)^n.
\end{eqnarray}

\section{Heralded spontaneous parametric down-conversion}

The photon source that forms the basis of our architecture is an SPDC source operating with a high-repetition rate (i.e. being rapidly pumped). An SPDC source comprises a crystal with a second-order nonlinearity, which is pumped with a strong coherent state. It prepares the two-mode state
\begin{eqnarray} \label{eq:SPDC_output}
\ket{\psi_\mathrm{src}} = \sum_{n=0}^\infty \lambda_n \ket{n,n},
\end{eqnarray}
and the photon-number distribution in each mode is thermally distributed:
\begin{eqnarray} \label{eq:SPDC_thermal_stat}
p_\mathrm{src}(n) = |\lambda_n|^2 = \frac{1}{\bar{n}+1} \left(\frac{\bar{n}}{\bar{n}+1}\right)^n.
\end{eqnarray}
Here $\bar{n}$ is the average photon-number, which in present-day experiments is intentionally kept very low so as to reduce the probability of multiple-pair creation events.

The commonly measured statistics for SPDC are Poissonian-distributed and this distribution is frequently used in the analysis of multiplexed sources \cite{14adam053834,13schmiegelow447,13mazzarella023848,11mower052326,bib:ZeilingerMultiplex,bib:ShapiroMultiplex,06peters630507,04jeffrey1}, often also with the probabilities of generating a single or multiple photons used as figures of merit. The Poisson distribution arises when the collected light includes many spatial and/or frequency modes since this is the limit of the convolution of many thermally distributed modes. If the output of the SPDC modes are carefully filtered it is possible to select a single mode yielding a photon number distribution with thermal statistics of Eq.~\ref{eq:SPDC_thermal_stat} \cite{12dovrat2266, 09mauerer053815}. 

In this work we are considering a source suitable for quantum information tasks so the appropriate model is one where a single photon is prepared in a well defined mode so that it is as indistinguishable from other such photons as possible. This will allow full non-classical interference in subsequent photonic circuits. The fidelity of the photon output with an ideal single-photon state is the appropriate figure of merit for these tasks. If many modes are collected in the output state, the result will be a statistical mixture and this will reduce the overall fidelity against the ideal. As noted in the introduction, that using a single apparatus to generate the photons also naturally helps in producing indistinguishable photons.

The key observation about Eq.~\ref{eq:SPDC_output} is that the photon-numbers in the two modes are perfectly correlated. Thus in principle with ideal detectors, if we detect some photon-number in the first mode, we can guarantee the same photon-number in the other mode. It is this correlation in photon-number that makes SPDC sources high-quality, reliable heralded sources when combined with high-efficiency detectors. The main problem with these sources is that the heralding protocol is inherently non-deterministic. Thus the success probability of obtaining $m$ single photons from $m$ sources for use in a larger protocol drops exponentially with $m$. This is the motivation for considering multiplexed schemes, as they can significantly boost the single-photon component at the output of a single device.

The photo-detection efficiency plays a major role in the fidelity of the prepared state. This is because detector inefficiency results in higher-order photon-number events being recorded as single-photon events, thereby preparing a state which is a mixture of the desired single-photon term, plus additional unwanted higher-order terms \cite{09barbieri209}. While the detrimental effect on fidelity can be reduced by setting \mbox{$\bar{n}\ll 1$}, and hence \mbox{$\left\vert\lambda_1\right\vert \ll 1$}, this impacts poorly on the efficiency. Moreover, imperfections such as dark counts and after-pulsing will further degrade the source.

Next we consider the problem of combining heralded single-photon state preparation with our model for an inefficient detector. If we take a single SPDC source, pulse it once, and post-select on the desired heralding outcome in the first mode, the reduced state in the unmeasured mode is given by,
\begin{eqnarray}
\hat\rho_\mathrm{resolved} &=&\mathcal{N}_r^{-1} \sum_{n=1}^\infty p_\mathrm{det}(1|n) p_\mathrm{src}(n) \ket{n}\bra{n}, \nonumber \\
\hat\rho_\mathrm{bucket} &=&\mathcal{N}_b^{-1} \sum_{n=1}^\infty p_\mathrm{det}(\mathrm{click}|n) p_\mathrm{src}(n) \ket{n}\bra{n},
\end{eqnarray}
in the cases of number-resolved and bucket detectors respectively. The normalisations are given by
\begin{align*}
\mathcal{N}_r &= \sum_{n=1}^\infty p_\mathrm{det}(1|n) p_\mathrm{src}(n) = \frac{\bar{n}\eta_d}{(1+\bar{n}\eta_d)^2}, \\
\mathcal{N}_b &= \sum_{n=1}^\infty p_\mathrm{det}(\mathrm{click}|n) p_\mathrm{src}(n) = \frac{\bar{n}\eta_d}{1+\bar{n}\eta_d}.
\end{align*}
Note that these states are perfect mixtures in the photon-number degree of freedom, with no coherences between different photon-numbers. 

The conditional probabilities of preparing $n$ photons given that heralding was successful, are 
\begin{eqnarray}
p_\mathrm{prep}(n|1) &=& \bra{n} \hat\rho_\mathrm{resolved} \ket{n} \nonumber \\
&=& p_\mathrm{det}(1|n) p_\mathrm{src}(n) /\mathcal{N}_r\nonumber \\
&=& \frac{n (\bar{n}-\eta_d \bar{n})^{n-1} (1+\eta_d \bar{n})^2}{(1+\bar{n})^{n+1}}, \nonumber \\
p_\mathrm{prep}(n|\mathrm{click}) &=& \bra{n} \hat\rho_\mathrm{bucket} \ket{n} \nonumber \\
&=& p_\mathrm{det}(\mathrm{click}|n) p_\mathrm{src}(n)/\mathcal{N}_b \nonumber \\
&=& \frac{\bar{n}^{n-1}(1-[1-\eta_d]^n)(1+\eta_d \bar{n})}{\eta_d(1+\bar{n})^{n+1}}.
\end{eqnarray}
For \mbox{$n=1$}, these are equivalent to the fidelities of single-photon state preparation, conditional upon a successful heralding event. Note that if we were using Poissonian distributed output for an SPDC this would not be the case.

The total probabilities of registering a successful heralding event are given by,
\begin{align} \label{eq:S}
\mathcal{S}_\mathrm{resolved} &= \sum_{n=1}^\infty p_\mathrm{det}(1|n) p_\mathrm{src}(n) = \mathcal{N}_r \nonumber \\
\mathcal{S}_\mathrm{bucket} &= \sum_{n=1}^\infty p_\mathrm{det}(\mathrm{click}|n) p_\mathrm{src}(n) = \mathcal{N}_b
\end{align}
for number-resolved and bucket detectors respectively. Note that \mbox{$\mathcal{S}_\mathrm{bucket}\geq \mathcal{S}_\mathrm{resolved}$}, since a bucket detector accepts more outcomes than a number-resolved detector.

When number-resolved detectors of perfect efficiency are employed, all of the events in $\mathcal{S}$ correspond to perfect single-photon state preparation. However, for bucket detectors or inefficient detectors, some of these events will correspond to having prepared the wrong photon-number.

\section{Temporal multiplexing}

To increase the efficiency of a single SPDC source we imagine pumping it $t$ times in succession, preparing a pulse-train of $t$ time-bins but using only a single time-bin as output. The total probabilities of measuring \emph{at least} one successful heralding event are,
\begin{eqnarray} \label{eq:St}
\mathcal{S}_\mathrm{resolved}(t) &=& 1 - (1-\mathcal{S}_\mathrm{resolved})^t \nonumber \\
&=& 1 - \left[1 - \frac{\eta_d \bar{n}}{(1+\eta_d \bar{n})^2}\right]^t, \nonumber \\
\mathcal{S}_\mathrm{bucket}(t) &=& 1 - (1-\mathcal{S}_\mathrm{bucket})^t \nonumber \\
&=& 1-\left[\frac{1}{1+\eta_d \bar{n}}\right]^t.
\end{eqnarray}

\begin{figure}[htb]
\centering
\includegraphics[width=0.9\columnwidth]{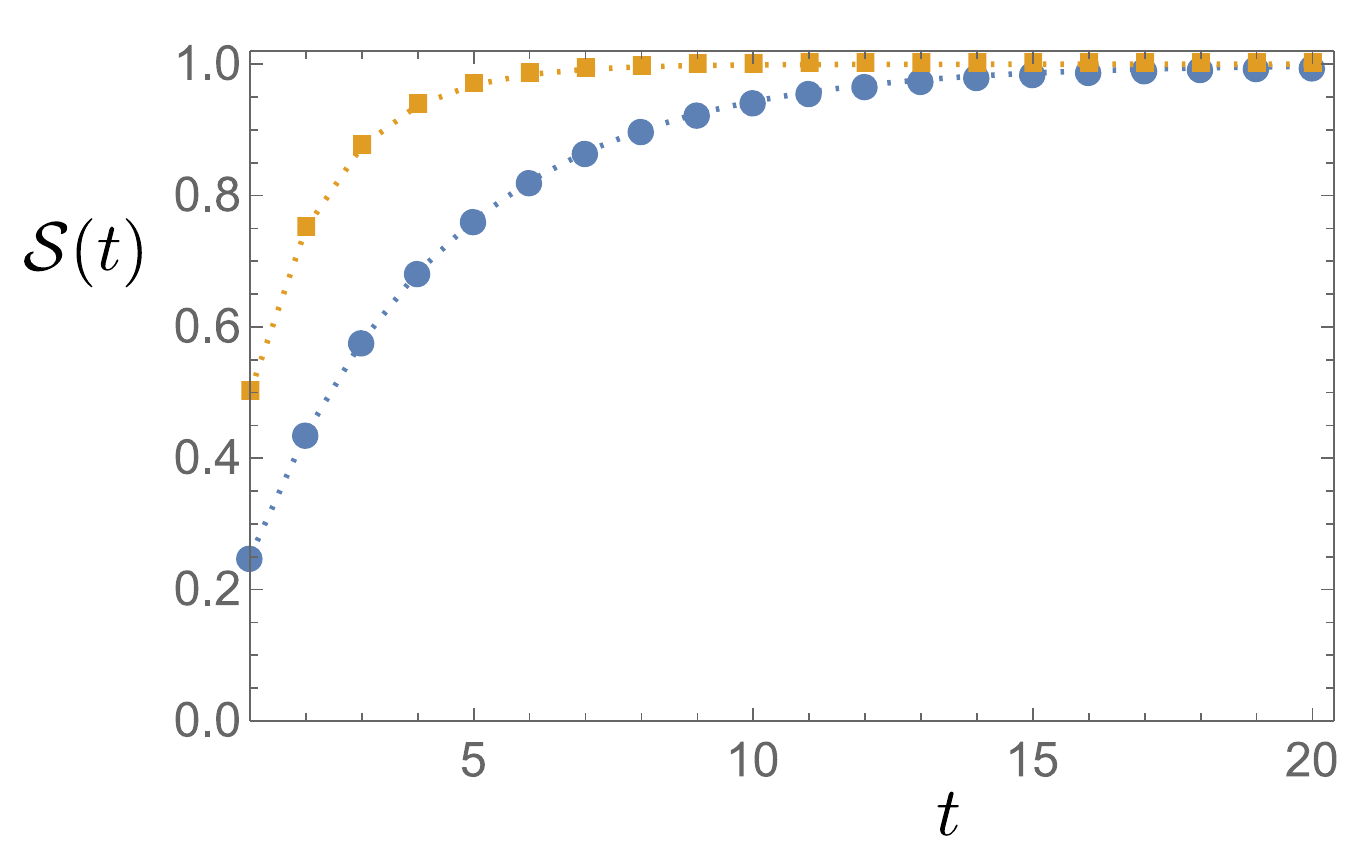}
\caption{Heralding probability (Eq.~\ref{eq:St}) against number of time-bins, for both number-resolved (blue circles) and bucket detectors (yellow squares), where \mbox{$\eta_d=1$} and \mbox{$\bar{n}=1$}.} \label{fig:ideal_S_vs_t}
\end{figure}

The heralding probabilities against the number of time-bins $t$ is shown in Fig.~\ref{fig:ideal_S_vs_t} for both number-resolved (blue) and bucket detectors (yellow), assuming perfect efficiency. Note that bucket detectors yield a heralding probability strictly greater than that of number-resolved detectors. As mentioned earlier this is simply because bucket detectors interpret any higher-order photon-number term as being a successful heralding event. Clearly such mistaken events would degrade the quality of the prepared output state so we need to consider the fidelity of the output as well as the probability.  We remark that in Eq.~\ref{eq:S} the mean photon number $\bar{n}$ and detector efficiency $\eta_d$ only appear as a product. Indeed, all expressions in the paper can rewritten so that the efficiency $\eta_d$ only ever appears in this combination, but for simplicity we leave them distinct.  

When combining the SPDC source with the fibre-loop architecture as shown in Fig.~\ref{fig:arch}, different time-bins in the heralded output mode may pass through the loop different numbers of times and thus suffer different amounts of loss. The net transmission of a time-bin subject to $l$ iterations through the loop is given by,
\begin{equation}
\tau_l = {\eta_s}^{l+1} {\eta_f}^l.
\end{equation}
That is, the time-bin passes through the switch once, and then through the loop and switch an additional time for every iteration of the loop.

Clearly there are many strategies that could be employed when controlling the switching sequence in the architecture. Most trivially, as soon as an SPDC heralding event succeeds, we could simply keep the associated heralded photon stored in the fibre-loop until we reach the last time-bin, at which point we switch it out. However, this would be sub-optimal in terms of state fidelity, since this photon, were it to occur at an early stage, would be subject to many traverses of both  the fibre-loop and the switch.

The optimal strategy in terms of minimising loss (and hence maximising fidelity), is to choose our switching sequence so as to keep the \emph{most recent} successfully heralded photon as our final output state.

This is achieved as follows. Whenever an SPDC heralding event succeeds, we couple the corresponding state completely into the fibre-loop, thereby coupling out what was previously in memory. We then toggle the switch so as to keep the state in memory. The state stays in memory unless another SPDC heralding event succeeds at a later stage, at which point we couple out the state presently in memory, and couple in the newly prepared state. This is repeated for the duration of the protocol.



To calculate the fidelities of the state in the last time-bin with a single photon state, we must sum over all the ways that only 1 photon reaches the output time-bin.  If the most recent time-bin corresponding to a successful heralding event passes through $l$ loops, and a heralding event corresponds to $n$ photons in this state before the switch, we find
\begin{gather}
\mathcal{F}_\mathrm{resolved}(l) = \sum_{n=1}^\infty p_\mathrm{prep}(n|1) \tau_l [1 - \tau_l]^{n-1} n, \nonumber \\
=\frac{\tau_l(1\!+\!\bar{n}\eta_d)^2[1+\bar{n}+\bar{n}(1\!-\!\eta_d)(1\!-\!\tau_l)]}{[1+\bar{n}-\bar{n}(1\!-\!\eta_d)(1\!-\!\tau_l)]^3},\nonumber \\
\mathcal{F}_\mathrm{bucket}(l) = \sum_{n=1}^\infty p_\mathrm{prep}(n|\mathrm{click}) \tau_l [1 - \tau_l]^{n-1} n, \nonumber \\
=\frac{\tau_l(1\!+\!\bar{n}\eta_d)[1+2\bar{n}+\bar{n}^2\eta_d+\bar{n}^2\tau_l(1\!-\!\eta_d)(2\!-\!\tau_l)]}{(1+\bar{n}\tau_l)^2(1\!+\!\bar{n}[(1\!-\!\eta_d)\tau_l+\eta_d])^2}.
\end{gather}
Here we have modelled the loss by mixing the output state with vacuum on a beamsplitter with a transmission $\tau_l$. Note that the fidelities take into account `accidental' single photon creation where a higher photon number is created due to heralding inefficiencies but then the right number of photons are lost to return the output state to a single photon state.

For a given measurement signature, these results allow us to calculate the exact fidelity associated with that signature. However, in reality we wish to accept \emph{all} heralding events that are successful, and not post-select upon particular heralding events, which would entirely defeat the purpose of multiplexing. If we accept \emph{any} heralding event that is successful, but only keep the most recent respective time-bin, the average fidelity of the prepared state is given by an average over all possible acceptable heralding signatures, weighted according to their prevalence.

Enumerating the events in decreasing order of desirability we label as $l\!=\!0$ the case where a photon is heralded in the \emph{last} time-bin. This has probability $\mathcal{S}$ given by the appropriate expression in Eq.~\ref{eq:S}. Note the photon does not go through the loop in this case. The next event $l\!=\!1$ sees a heralded photon appear in the penultimate time-bin and no photon in the last (or we would have accepted that event) and so has probability $\mathcal{S}(1-\mathcal{S})$. Labelling the time-bins in reverse chronological order from zero gives a probability of $\mathcal{S}(1-\mathcal{S})^l$ that we will accept a photon heralded in time-bin $l$ which passes through the loop $l$ times. Finally we append the last possible event where no photon is heralded in the entire train $t$ which occurs with probability $(1-\mathcal{S})^t$. This forms a probability distribution for the resulting events of the switching strategy given by,
\begin{equation}\label{eq:probaiities}
p(l) = \begin{cases}
\mathcal{S}(1-\mathcal{S})^l  &0 \le l < t\\
(1-\mathcal{S})^l  &l=t
\end{cases}.
\end{equation}

Our goal is to determine when the source will yield an output photon and what the quality of that photon is. So for calculating the average fidelities we exclude the case where no heralding event occurred ($l=t$). This is permissible as we have a record of when the source was successful through the heralding signature. Excluding the last event means we need to normalise the distribution by $1\!-\!p(t)=1\!-\!(1\!-\!\mathcal{S})^t$. Thus the conditional average fidelities are given by,
\begin{align}
\label{eq:Fcond}\Fcond_\mathrm{resolved} 
&= \frac{\mathcal{S}_\mathrm{resolved}}{1\!-\!(1\!-\!\mathcal{S}_\mathrm{resolved})^t}
\sum_{l=0}^{t-1} \mathcal{F}_\mathrm{resolved}(l) [1\!-\!\mathcal{S}_\mathrm{resolved}]^l, \nonumber \\
\Fcond_\mathrm{bucket} 
&=  \frac{\mathcal{S}_\mathrm{bucket}}{1\!-\!(1\!-\!\mathcal{S}_\mathrm{bucket})^t}
\sum_{l=0}^{t-1} \mathcal{F}_\mathrm{bucket}(l) [1-\mathcal{S}_\mathrm{bucket}]^l.
\end{align}

In the case of perfect efficiency through the system (\mbox{$\eta_s=\eta_f=\eta_d=1$}), these fidelities and heralding probabilities reduce to,
\begin{eqnarray}
\Fcond_\mathrm{resolved} &=& 1, \nonumber \\
\mathcal{S}_\mathrm{resolved} &=& 1 - \left[1 - \frac{\bar{n}}{(1+\bar{n})^2}\right]^t, \nonumber \\
\Fcond_\mathrm{bucket} &=& \frac{1}{1+\bar{n}}, \nonumber \\
\mathcal{S}_\mathrm{bucket} &=& 1-\left[\frac{1}{1+\bar{n}}\right]^t.
\end{eqnarray}
In that case, the fidelities are independent of $t$ since all time-bins are optimal and the multiplexing sequence is irrelevant. With number-resolved detectors the fidelity is perfect, irrespective of $\bar{n}$. Thus, the optimal strategy is simply to use as many time-bins as possible so as to maximise the heralding efficiency, as pointed out in \cite{bib:ChristMultiplex}. But for bucket detectors the fidelity is optimised in the limit of \mbox{$\bar{n}\to 0$}. Of course, in this limit the heralding probability vanishes. Thus, with bucket detectors there is a trade-off between fidelity and heralding probability (one very familiar to experimentalists developing heralded photon sources).

\begin{figure}[!htb]
\includegraphics[width=\columnwidth]{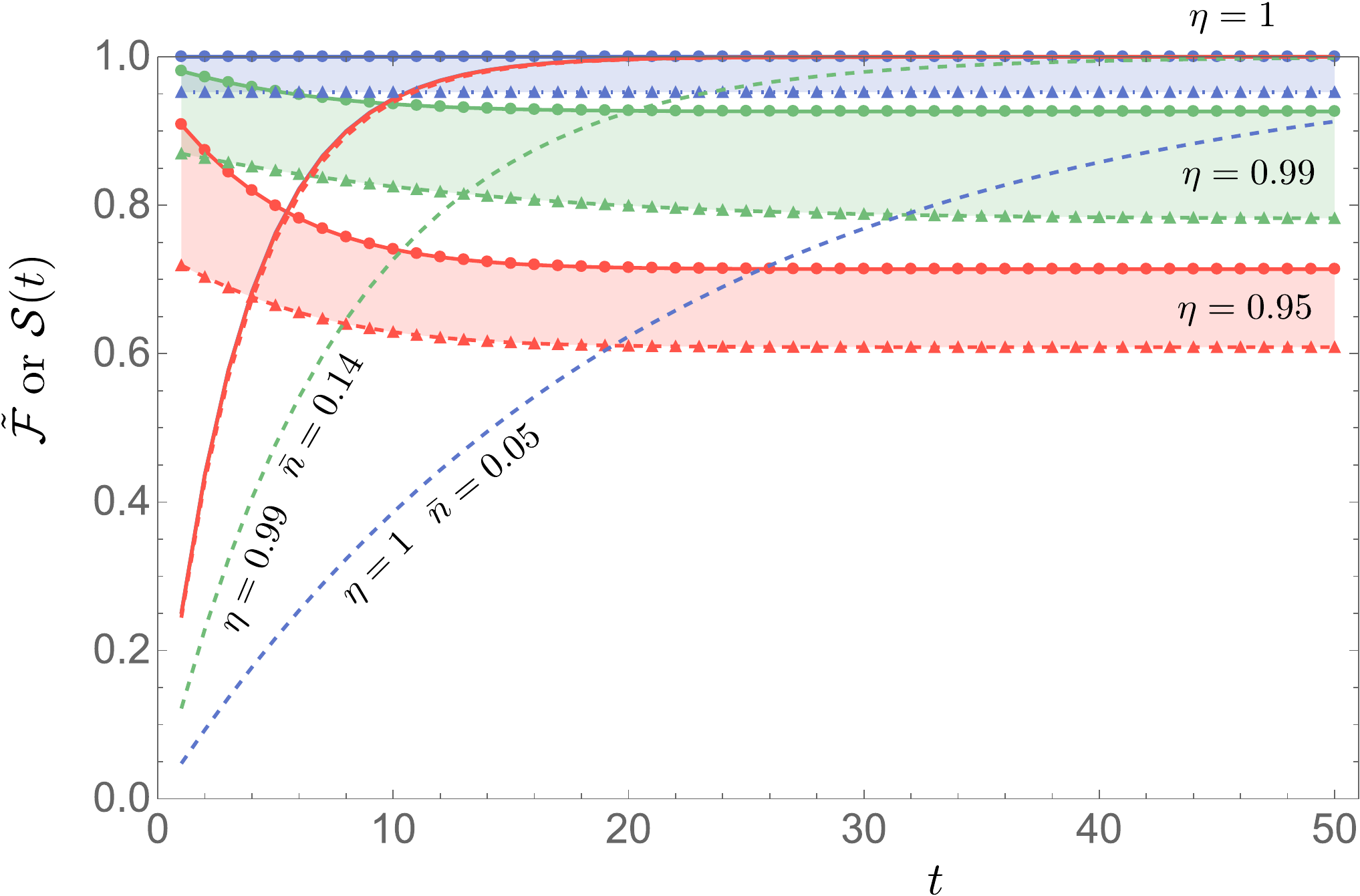}
\caption{Fidelity and heralding probabilities for various sources for which all specific efficiencies are the same: $\eta=\eta_s=\eta_f=\eta_d$. Decreasing curves plot average conditional fidelities $\Fcond_\mathrm{resolved}$ (circles and solid line) and $\Fcond_\mathrm{bucket}$ (triangles and dashed line) for reducing efficiencies from top to bottom of $\eta=1.0$ (blue), $\eta=0.99$ (green), and $\eta=0.95$ (red).  Increasing curves plot $\mathcal{S}_\mathrm{resolved}(t)$ (solid) and $\mathcal{S}_\mathrm{bucket}(t)$ (dashed) for same efficiencies. The shaded bands indicate the improvement possible by detector type, the bands themselves indicate improvement by increasing efficiency. The optimum $\bar{n}$ for achieving the highest $\Fcond$ at $t=50$ was chosen for each non-unit efficiency detector. These were $\bar{n}=0.90$ and $\bar{n}=0.95$ for the resolved detector with efficiencies of 0.99 and 0.95, and $\bar{n}=0.14$ and $\bar{n}=0.34$ for the bucket detector with the same efficiencies respectively. For the unit-efficiency curves $\bar{n}$ was set to 0.05 for the bucket detector (which is actually maximised for $\bar{n}\rightarrow 0$) and 1.0 for the resolved detector which maximises $\mathcal{S}_\mathrm{resolved}$ instead. } 
\label{fig:SF_vs_t}
\end{figure}

In Fig.~\ref{fig:SF_vs_t} we show the conditional average fidelity and heralding probabilities for different efficiencies and the two detectors where the source $\bar{n}$ has been optimised for the highest fidelity. The figure captures the effect of several parameters on the quality of the source. There is an overall trade-off between heralding probability and fidelity. As the number of time-bins increases, the heralding probability asymptotically approaches unity, but at the same time the average fidelity decreases. The improvement in the fidelity by using a heralded detector over a bucket detector is lost as the number of time-bins increases, but the main factor affecting the quality of the output is the efficiency of the heralding detector.

\begin{figure}[htb]
\includegraphics[width=\columnwidth]{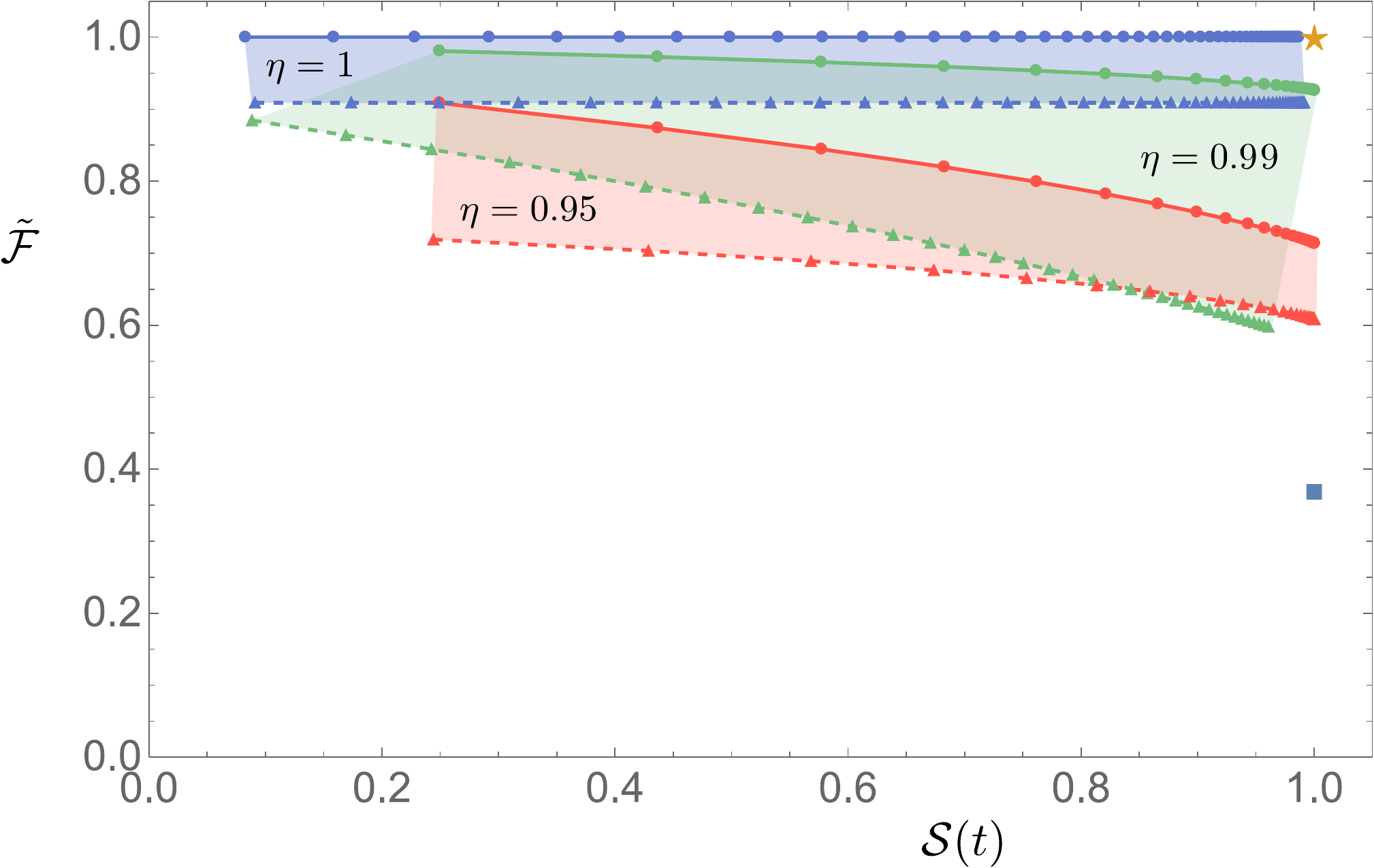}
\caption{Conditional average fidelity versus heralding probability, parametrised by the number of time-bins (increasing from left to right), using sources with $\bar{n}$ set as described in Fig.~\ref{fig:SF_vs_t}. In each band the number-resolved detectors are shown by circles and solid line, and the bucket detectors by triangles and dashed line. The different bands represent different efficiencies from top to bottom of \mbox{$\eta=1.0$} (blue), \mbox{$\eta=0.99$} (green), and \mbox{$\eta=0.95$} (red), where all specific efficiencies are equal (\mbox{$\eta=\eta_s=\eta_f=\eta_d$}). More deterministic  sources lie to the right, higher quality sources lie to the top. For comparison, the star in the top-right of the plot is an ideal deterministic single photon source and the square below it is a coherent state $\ket{\alpha}$ with $|\alpha|^2=1$ which is sometimes used as a substitute.  } \label{fig:F_vs_S}
\end{figure}

The trade-off between heralding probability and fidelity can be seen even more directly if we plot one against the other as in Fig.~\ref{fig:F_vs_S}, where we use the number of time-bins as the parametrisation variable. Generally the closer to the top-right of the plot the better the source.

\begin{figure}[htb]
\centering
\includegraphics[width=0.9\columnwidth]{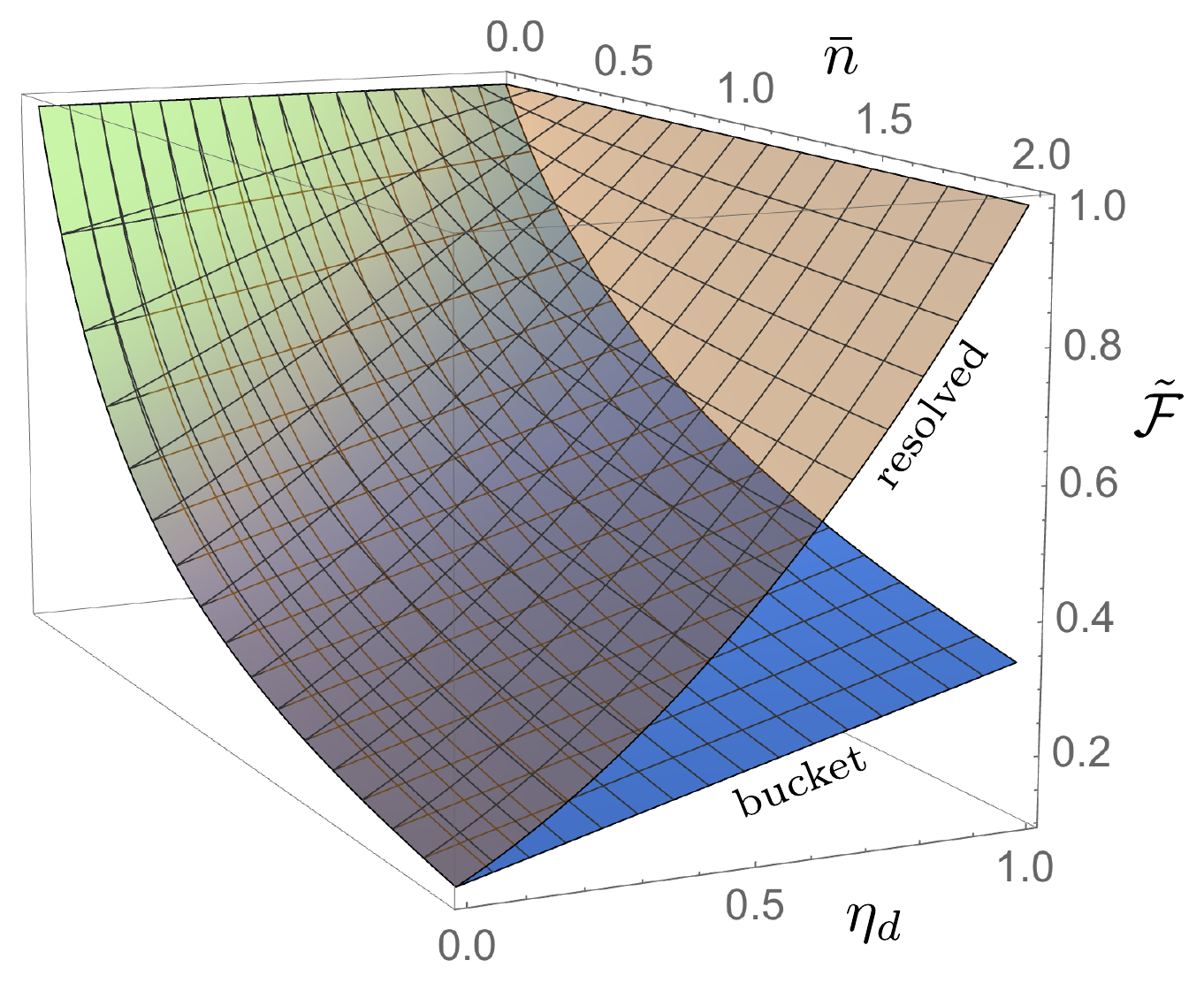} 
\caption{Average conditional fidelity $\Fcond$ against detector efficiency $\eta_d$ and SPDC mean photon-number $\bar{n}$, where the fibre-loop and switch are assumed to have perfect efficiency, \mbox{$\eta_f=\eta_s=1$}. (translucent yellow surface) Number-resolved detectors, (solid blue surface) bucket detectors. Note that in this case the fidelities are independent of the number of time-bins, since iterations through the loop do not degrade the stored state.} \label{fig:F_3D}
\end{figure}

In future integrated optical systems, it is plausible that integrated switches and delay lines might be highly efficient. In this instance the fidelity of the system will be limited by detector efficiency. Making the assumption that fibre-loops and switches have perfect efficiency, the fidelity of the prepared states are given by,
\begin{eqnarray}
\Fcond_\mathrm{resolved} &=& \left[\frac{1+\eta_d\bar{n}}{1+\bar{n}}\right]^2, \nonumber \\
\Fcond_\mathrm{bucket} &=& \frac{1+\eta_d\bar{n}}{(1+\bar{n})^2}.
\end{eqnarray}
These are shown in Fig.~\ref{fig:F_3D}. Note that,
\begin{eqnarray}
\mathcal{S}_\mathrm{resolved} &\leq& \mathcal{S}_\mathrm{bucket}, \nonumber \\
\Fcond_\mathrm{resolved} &\geq& \Fcond_\mathrm{bucket},
\end{eqnarray}
in all regimes. For either detector type we observe perfect fidelity when \mbox{$\bar{n}\to 0$}. However, the number-resolved detectors also yield perfect fidelity for any $\bar{n}$ if \mbox{$\eta_d=1$}. This is not the case when using bucket detectors, since even a perfectly efficient bucket detector will record higher order photon-number terms as containing a single photon.


\section{Black Box Operation}

\begin{figure*}[!htb]
\centering
\includegraphics[width=0.9\columnwidth]{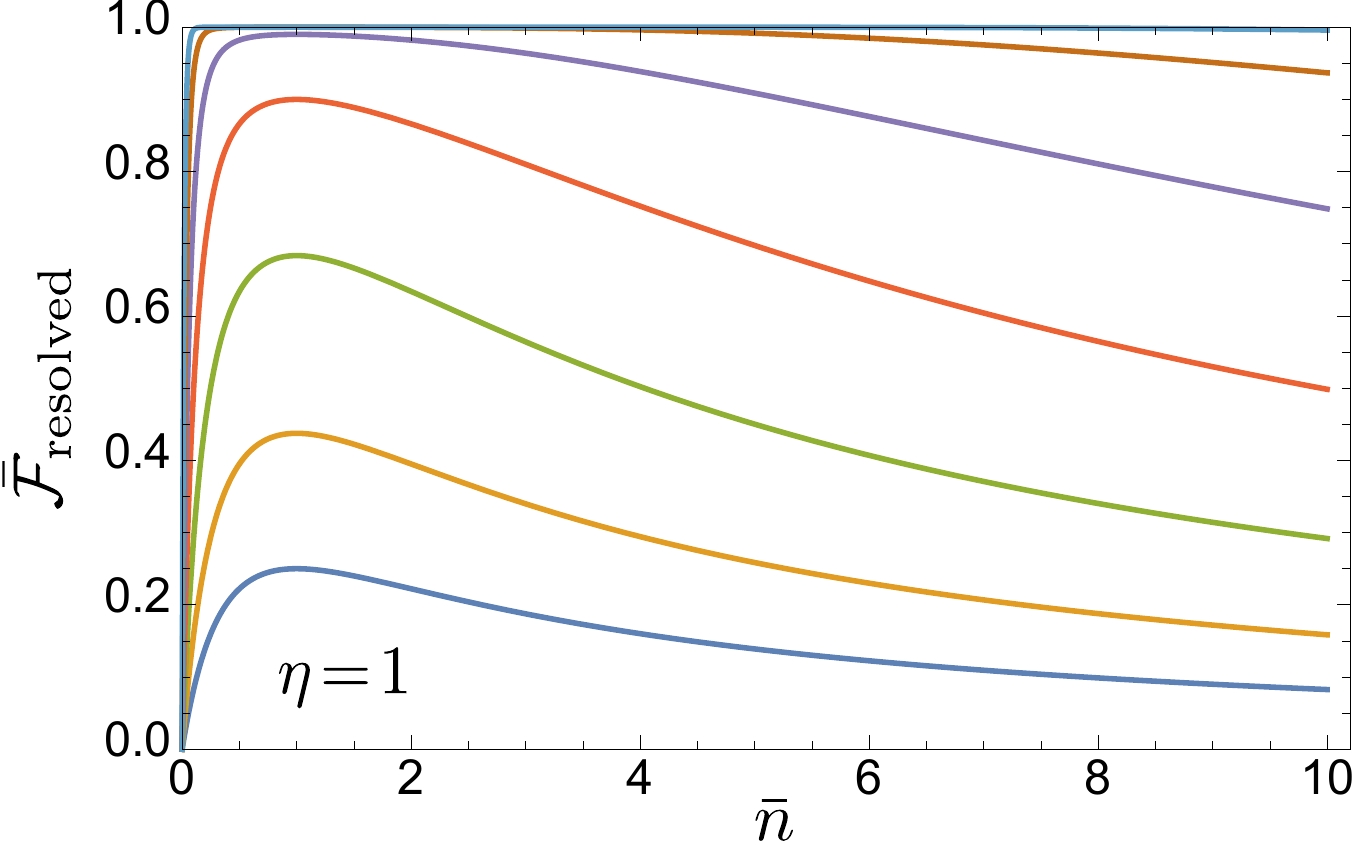}\qquad
\includegraphics[width=0.9\columnwidth]{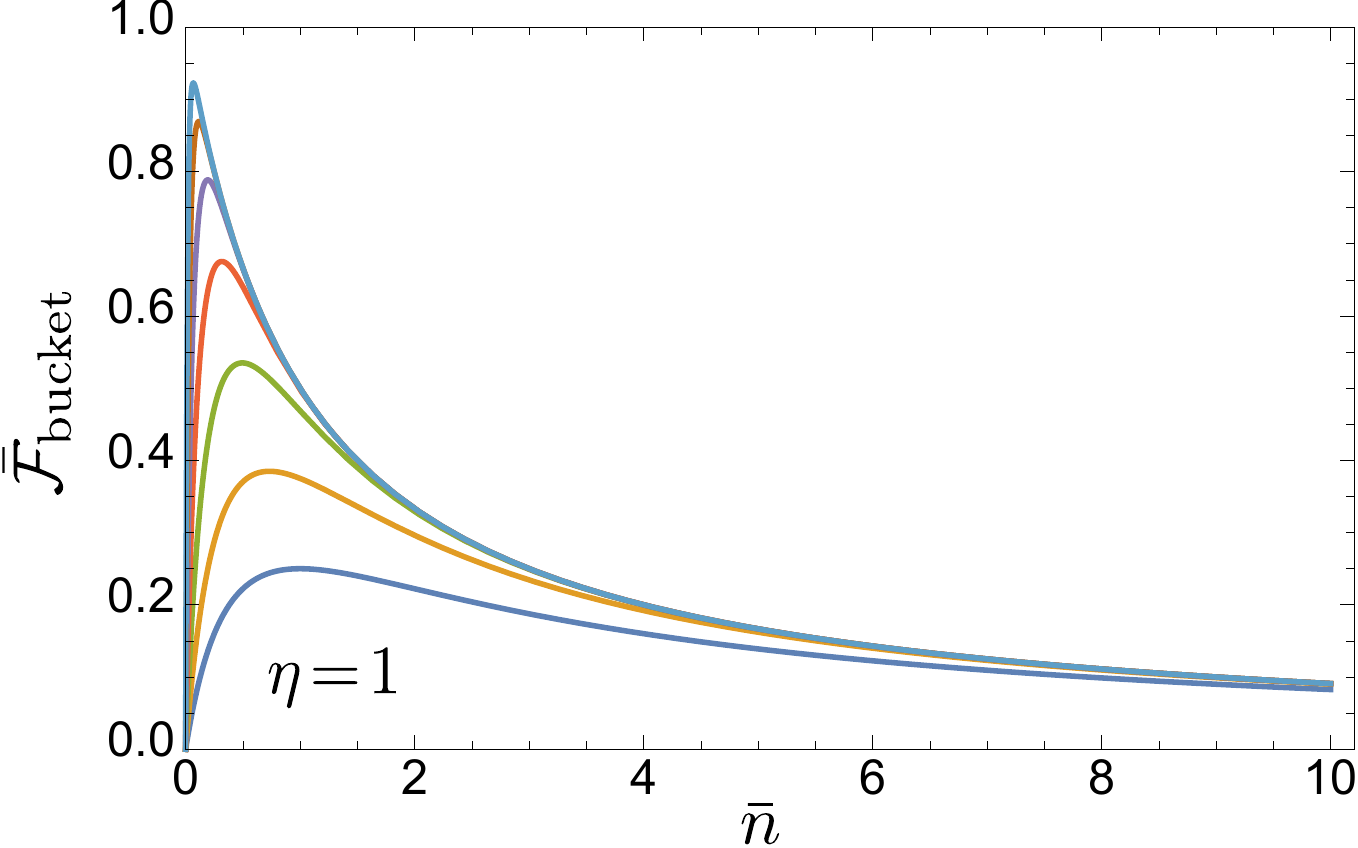}\\
\includegraphics[width=0.9\columnwidth]{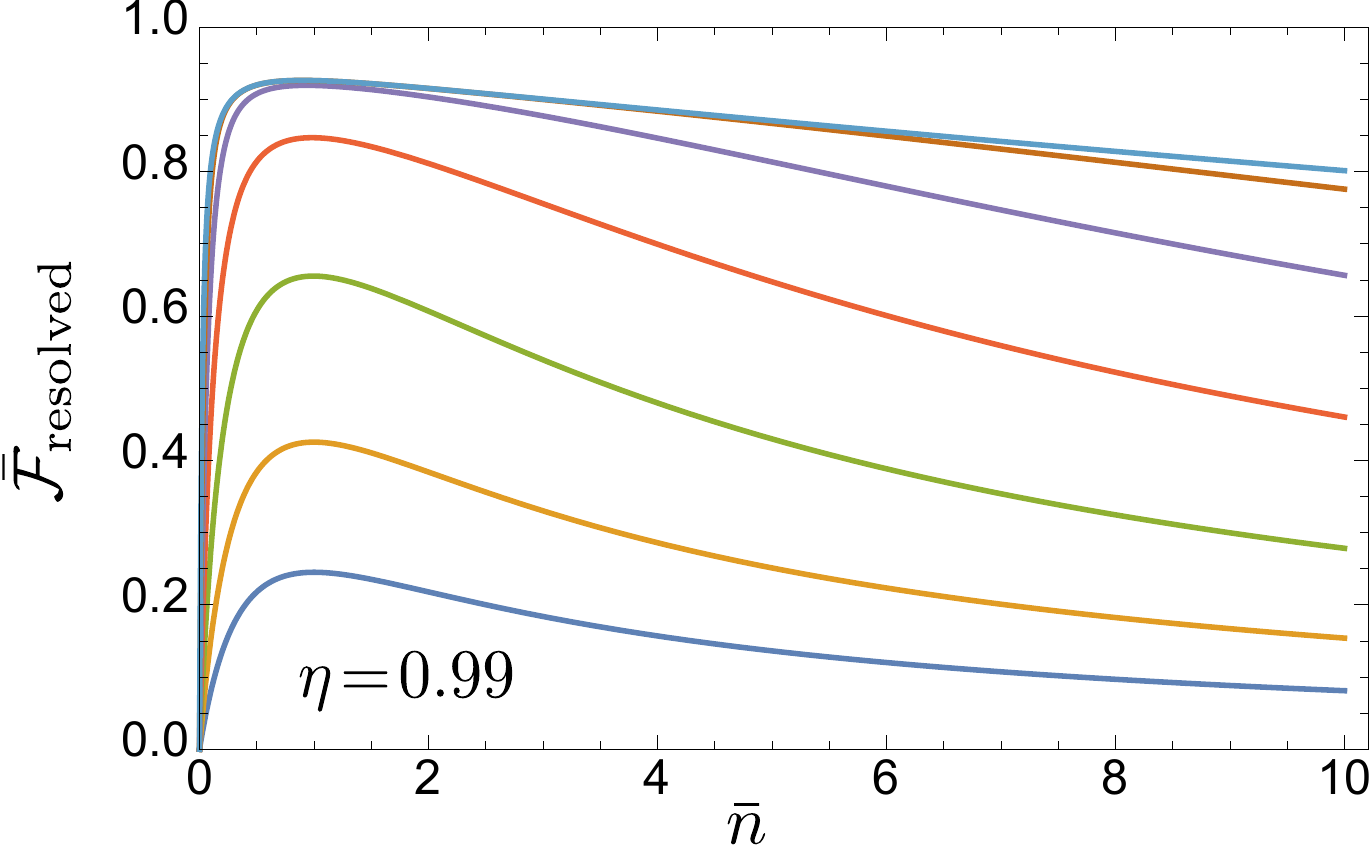}\qquad
\includegraphics[width=0.9\columnwidth]{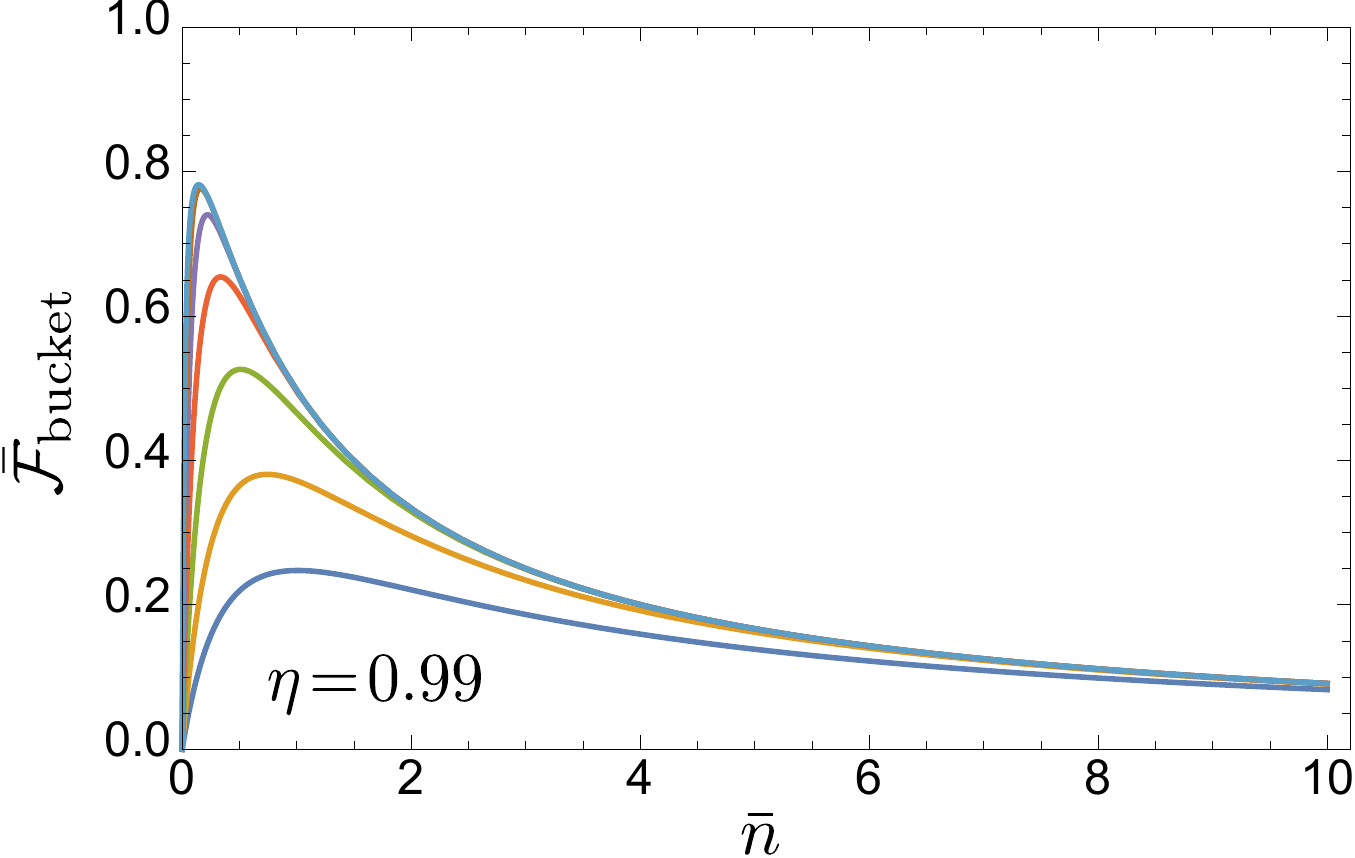}\\
\includegraphics[width=0.9\columnwidth]{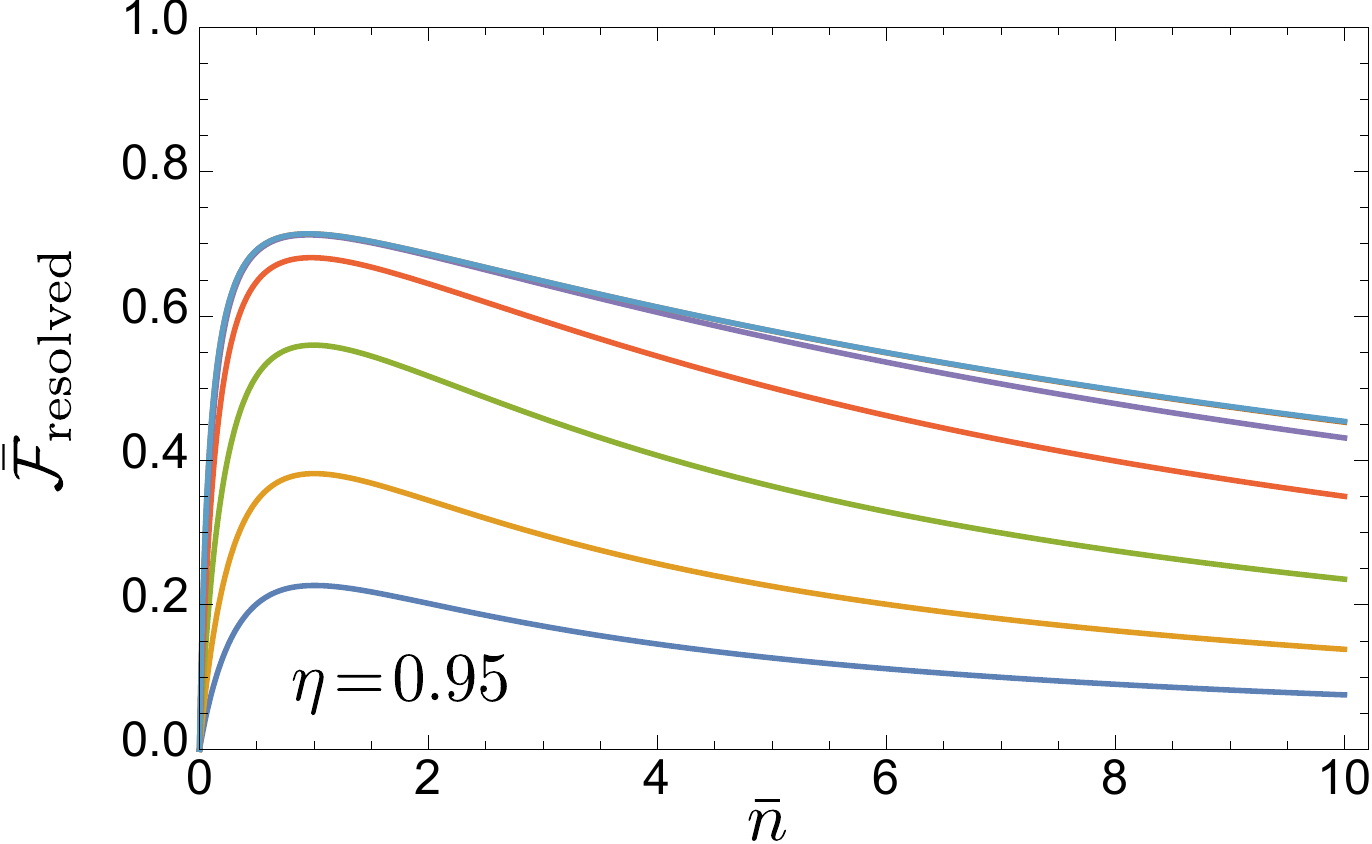}\qquad
\includegraphics[width=0.9\columnwidth]{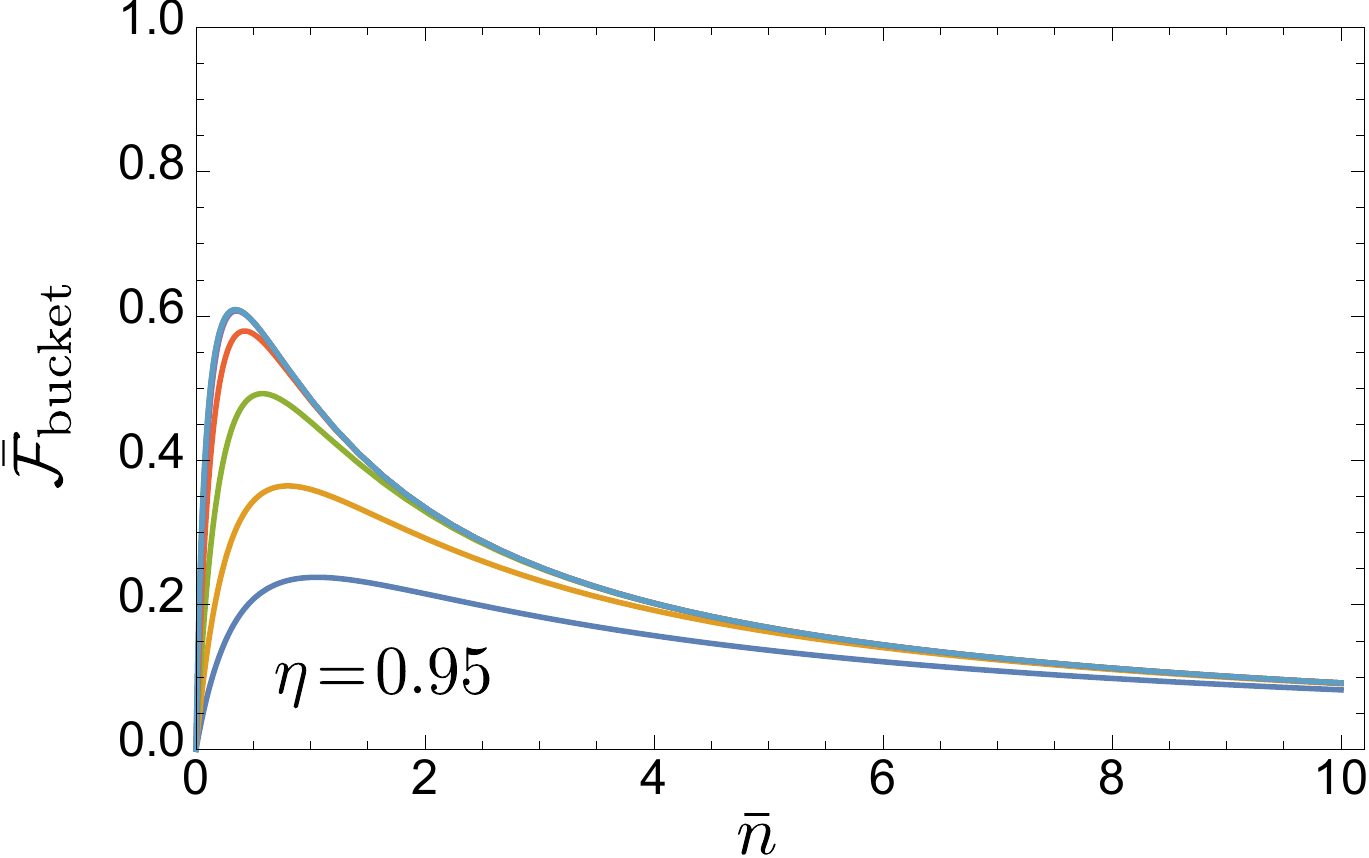}
\caption{Unconditional average fidelity, $\Funco$, as a function of \mbox{$\bar{n}$} for efficiencies \mbox{$\eta=1.0$} (top), \mbox{$\eta=0.99$} (middle), and \mbox{$\eta=0.95$} (bottom), with number-resolved (left) and bucket (right) detectors. In each plot, the curves from bottom to top are for an increasing number of time-bins in the sequence $t\in\{1,2,4,8,16,32,64\}$. All efficiencies are set equal, $\eta=\eta_s=\eta_f=\eta_d$}. \label{fig:Funcon}
\end{figure*}

So far we have explored using the loop architecture to make a non-deterministic heralded SPDC source more deterministic. The source is pumped $t$ times producing a train of $t$ time-bins each potentially containing heralded photons, and switching is used to increase the probability that one of those photons will be present in the last time-bin in the train. We demonstrated that this protocol has significant benefits in terms of heralding efficiency at the expense of some loss of fidelity. However the resulting source is still non-deterministic in the sense that the entire train of time-bins may fail to have a photon.

The device can also be operated as a `black box', whereby the user waits a set time and simply takes the output as is, regardless of any heralding signature, and we can determine the quality of this resulting source. Now the figure of merit to use is the unconditional average fidelity $\Funco$ over the probabilities in Eq.~\ref{eq:probaiities} that also include the no-photon outcome:
\begin{align}
\Funco_\mathrm{resolved} 
&= \mathcal{S}_\mathrm{resolved}
\sum_{j=0}^{t-1} \mathcal{F}_\mathrm{resolved}(j) [1\!-\!\mathcal{S}_\mathrm{resolved}]^j, \nonumber \\
\Funco_\mathrm{bucket} 
&= \mathcal{S}_\mathrm{bucket}
\sum_{j=0}^{t-1} \mathcal{F}_\mathrm{bucket}(j) [1-\mathcal{S}_\mathrm{bucket}]^j.
\end{align}
In the above we've made use of the fact that $\mathcal{F}=0$ when no photon is heralded in the entire train. These expressions are equivalent to the product of Eqs.~\ref{eq:St} and \ref{eq:Fcond}: $\Funco=\mathcal{S}(t)\Fcond$.

In Fig.~\ref{fig:Funcon} we plot $\Funco$ as a function of $\bar{n}$, for fixed efficiencies (\mbox{$\eta=\eta_s=\eta_f=\eta_d$}) and a representative total number of time-bins $t$ equal to 1, 2, 4, 8, 16, 32 or 64. For the efficiencies that we consider, it is seen that $\Funco$ increases as $t$ increases, and that there is an optimal $\bar{n}$ that decreases as $t$ increases. The shift to lower average photon-number as the number of time-bins increases helps to ensure that, for imperfect efficiencies, each heralding success corresponds to a large fidelity. For the relatively high efficiencies we consider, the interplay between the maximum unconditional fidelity, $\bar{n}$, and $\eta$ is more complex for bucket detectors than for number-resolved detectors, as heralding events that correspond to more than one photon can be balanced by losing all but one photon on the way to the output. Nevertheless, for large $\bar{n}$ the fidelity reduces to $\Funco_\mathrm{bucket}\to \eta^2/\bar{n}$ for all $t$.

In some situations it may be the case that the rate that photons are required is much lower than the repetition rate of the protocol. Alternatively we can imagine a bank of such sources that are left operating for a certain period until a large number of simultaneous photons are required. Both of these cases correspond to the large $t$ limit of the source. Since the heralding probability asymptotes to unity with increasing $t$, in this limit the source behaves as a true `push-button' source, releasing a photon on-demand. Figure~\ref{fig:SF_vs_t} suggests that in this limit the fidelity of the prepared state does not degrade to zero, but rather asymptotes to a value that is a function of the various efficiencies.

Of course, one would never actually leave the device running in the limit of \mbox{$t\to \infty$}, since this would contradict the notion of `on demand' state preparation, but as is evident from Fig.~\ref{fig:SF_vs_t}, more modest run times on the order of \mbox{$t\approx 30$} (in the case of \mbox{$\eta>0.95$}) are sufficient to bring us very close to the asymptotic state. An alternative arrangement would be to have a bank of such sources each primed to release a single photon on the push of a button. In this arrangement the duration of the transient is not so important as the simultaneous release and indistinguishability of all the photons.

\begin{figure}[htb]
\centering
\includegraphics[width=0.9\columnwidth]{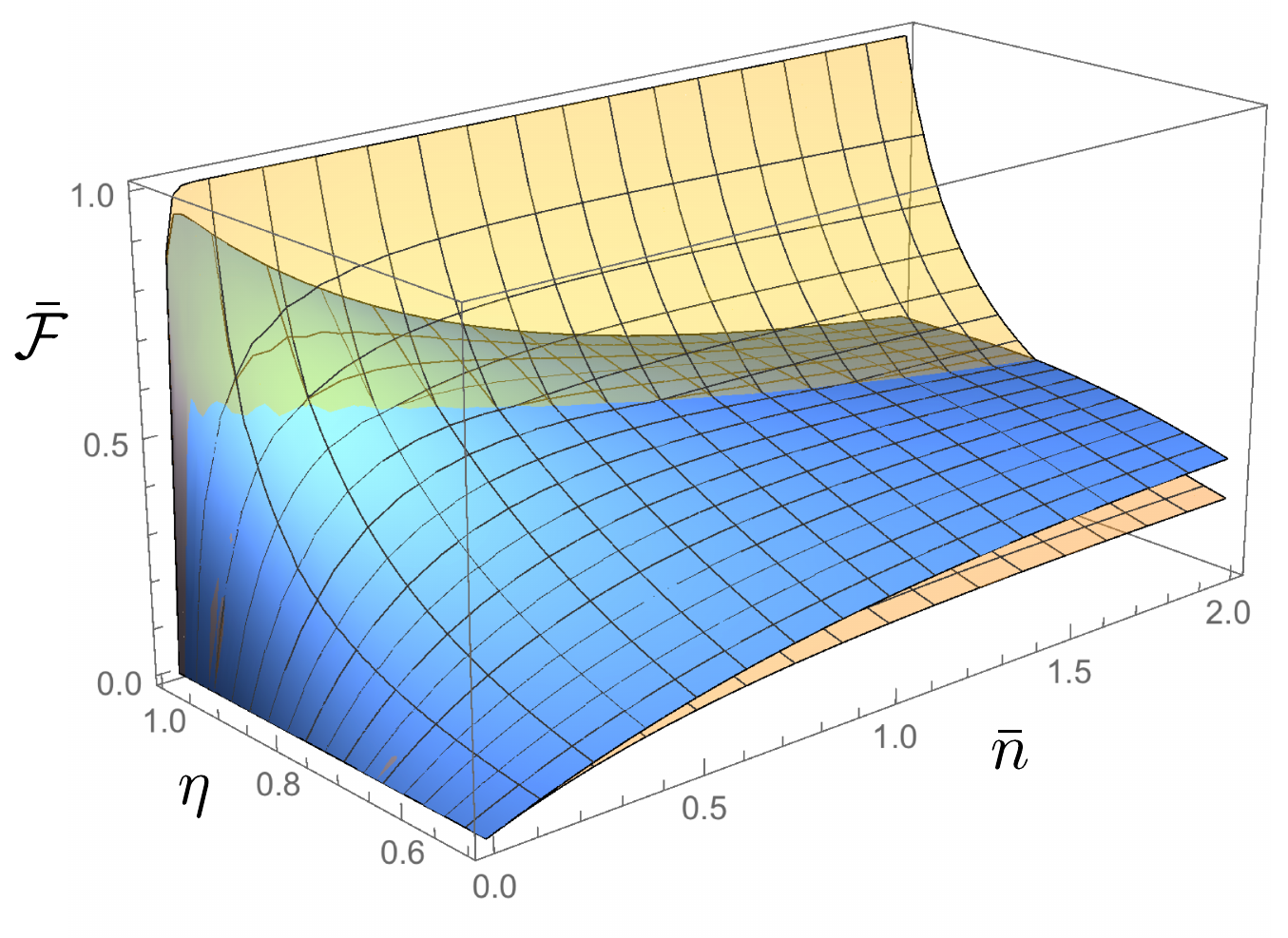} 
\caption{Long time limiting behaviour (with $t=100$) of the unconditional average fidelities, against $\bar{n}$ and \mbox{$\eta = \eta_f = \eta_s = \eta_d$}, where all efficiencies are equal. The translucent yellow surface is $\Funco_\mathrm{resolved}$, the opaque blue surface is $\Funco_\mathrm{bucket}$.} \label{fig:Finf}
\end{figure}

Figure~\ref{fig:Finf} shows this large $t$ behaviour for $t=100$ against $\bar{n}$ and $\eta$, where all efficiencies are equal (\mbox{$\eta=\eta_s=\eta_f=\eta_d$}). Thus, Fig.~\ref{fig:Finf} gives us the fidelities of a prepared state when the device is left running sufficiently long and the state in memory is coupled out on-demand.  Observe that for low efficiencies, the bucket detector fidelity (yellow) out-performs the number-resolved case (blue).  Again, this is because multi-pair events in which only one photon survives are able to contribute to the fidelity in the bucket case.

\section{Further Improvements}

Finally, we note some further improvements that could be made to the source. In Fig.~\ref{fig:SF_vs_t} and Fig.~\ref{fig:F_vs_S} we optimised $\bar{n}$ for the conditional average fidelity at a long-time limit. This strategy can be extended to optimising each pulse separately --- the first pulse is optimised for a photon heralded in the $(t-1)^\mathrm{th}$ time-bin, through to the last pulse optimised for the $0^\mathrm{th}$ time-bin. This could be accomplished by placing an additional dynamically controlled beamsplitter in the pump beam.

\begin{figure}[!htb]
\centering
\includegraphics[width=0.9\columnwidth]{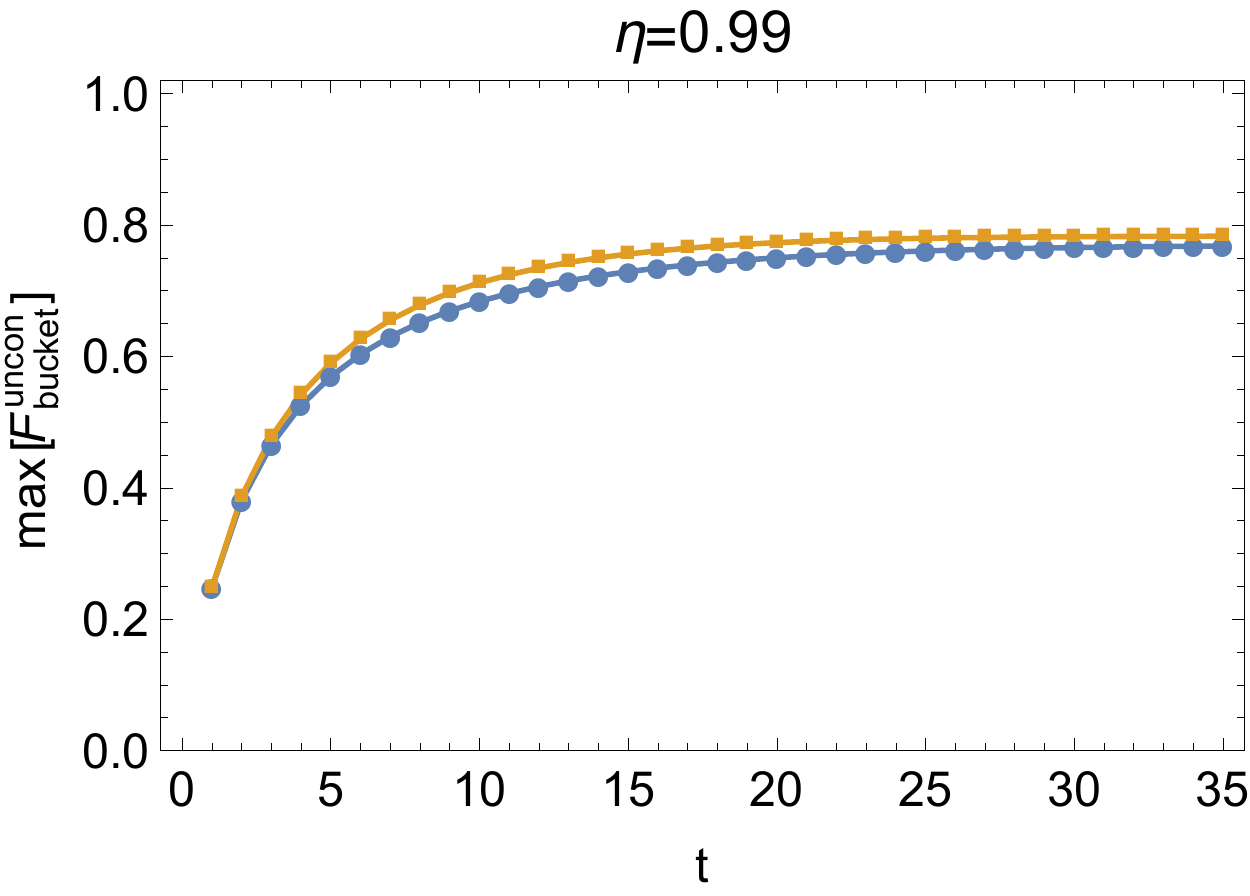}\\
\includegraphics[width=0.9\columnwidth]{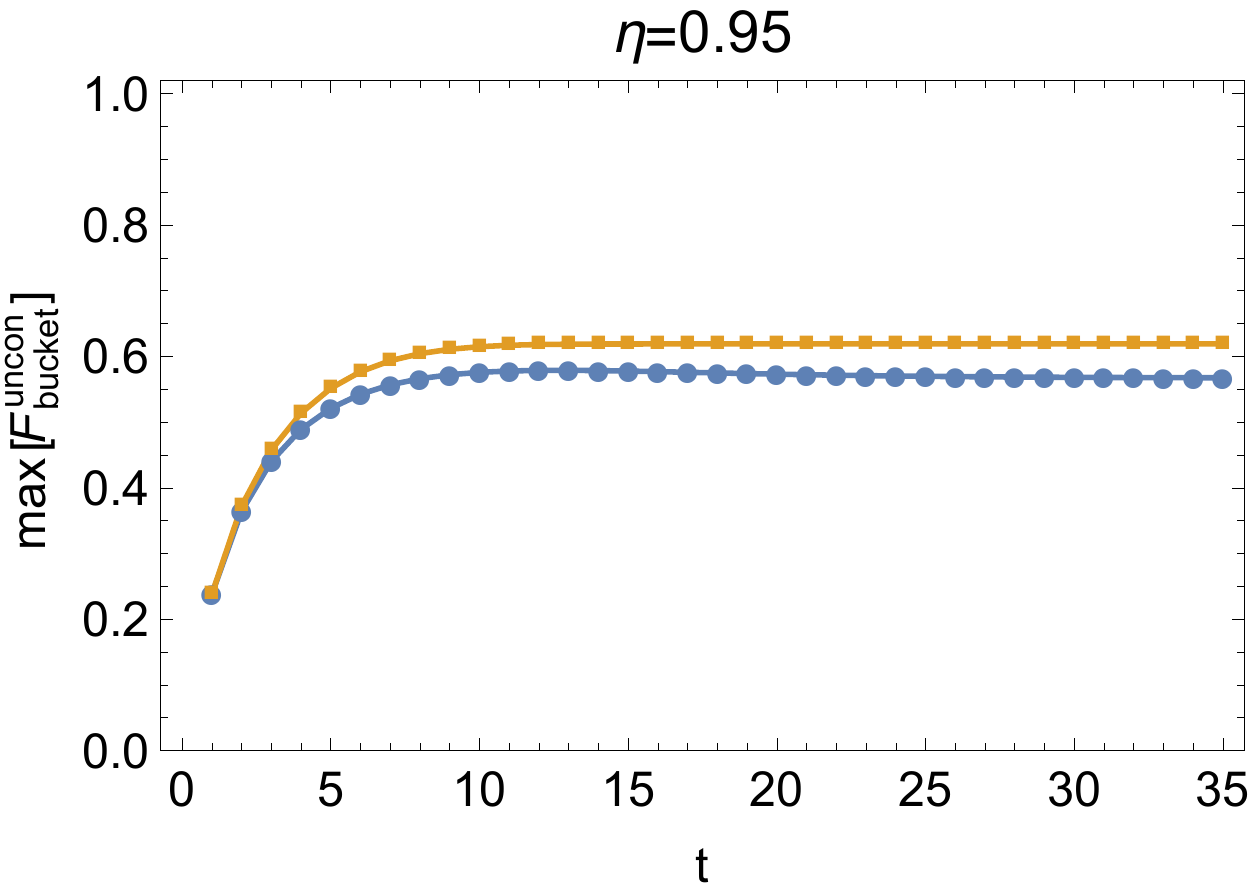}
\caption{Maximum unconditional average fidelity,  \mbox{$\max\left[\Funco\right]$}, as a function of the total number of time bins \mbox{$t$}, for unbiased (blue) and biased (yellow) operation given bucket detectors. (top) \mbox{$\eta=0.99$}. (bottom) \mbox{$\eta=0.95$}.} \label{fig:maxFS_unconditional}
\end{figure}

As an example, consider \mbox{$\eta_s=\eta_f=\eta_d=0.95$} and \mbox{$t=3$}. For these values, we calculate 
that $\Funco_\mathrm{bucket}$ reaches a maximum of $0.441$ for the constant \mbox{$\bar{n}=0.668$}, while for three different pump pulses with \mbox{$\bar{n}_1=1.31$}, \mbox{$\bar{n}_2=0.693$}, \mbox{$\bar{n}_3=0.466$}, \mbox{$\Funco_\mathrm{bucket}=0.458$}. In Fig.~\ref{fig:maxFS_unconditional} we plot the maximum unconditional fidelity achievable with constant operation, as well as this biased operation, as the number of time bins increases for fixed efficiencies with bucket detectors. This optimisation confers a larger advantage to bucket detectors than number-resolving detectors.

More significantly, aside from the black box operation the source carries a classical record of `quality' for every output time-bin --- we know which time-bin in the train gave a successful heralding event. The later in the time train that the photon was heralded, the less loss it has suffered.  We can then imagine having a bank of such sources and applying a spatial multiplexing scheme to pick the `freshest' photons in each time-bin. 

For two parallel sources, where we pick the source that produces the latest heralded time-bin as the output, it is possible to give the joint probability distribution under this strategy in analytical form. The probability that a photon was heralded in time-bin $j$ and not later is $\mathcal{S}(1\!-\!\mathcal{S})^j$. This event has to occur in at least one source, so there is one joint-event where both sources produce such photons and two lots of joint-events where only one source does and the other source produces an earlier time-bin photon, or non at all. All in all, the probability that the strategy produces a photon from time-bin $j$ is
\begin{gather}
\mathcal{S}(1\!-\!\mathcal{S})^j\left[ \mathcal{S}(1\!-\!\mathcal{S})^j + 2\!\!\sum_{k=j+1}^{t-1}\mathcal{S}(1\!-\!\mathcal{S})^k+2(1\!-\!\mathcal{S})^t \right]\nonumber\\
= \mathcal{S}(1-\mathcal{S})^{2j}(2-\mathcal{S}).
\end{gather}
After including the joint event where both sources fail to produce photons, the probability distribution of events for the two sources is
\begin{equation}\label{eq:probaiities2}
p_2(j) = \begin{cases}
\mathcal{S}(1-\mathcal{S})^{2j}(2-\mathcal{S})  &0 \le j < t\\
(1-\mathcal{S})^{2j}  &j=t
\end{cases}.
\end{equation}

\begin{figure}[htb]
\centering
\includegraphics[width=0.9\columnwidth]{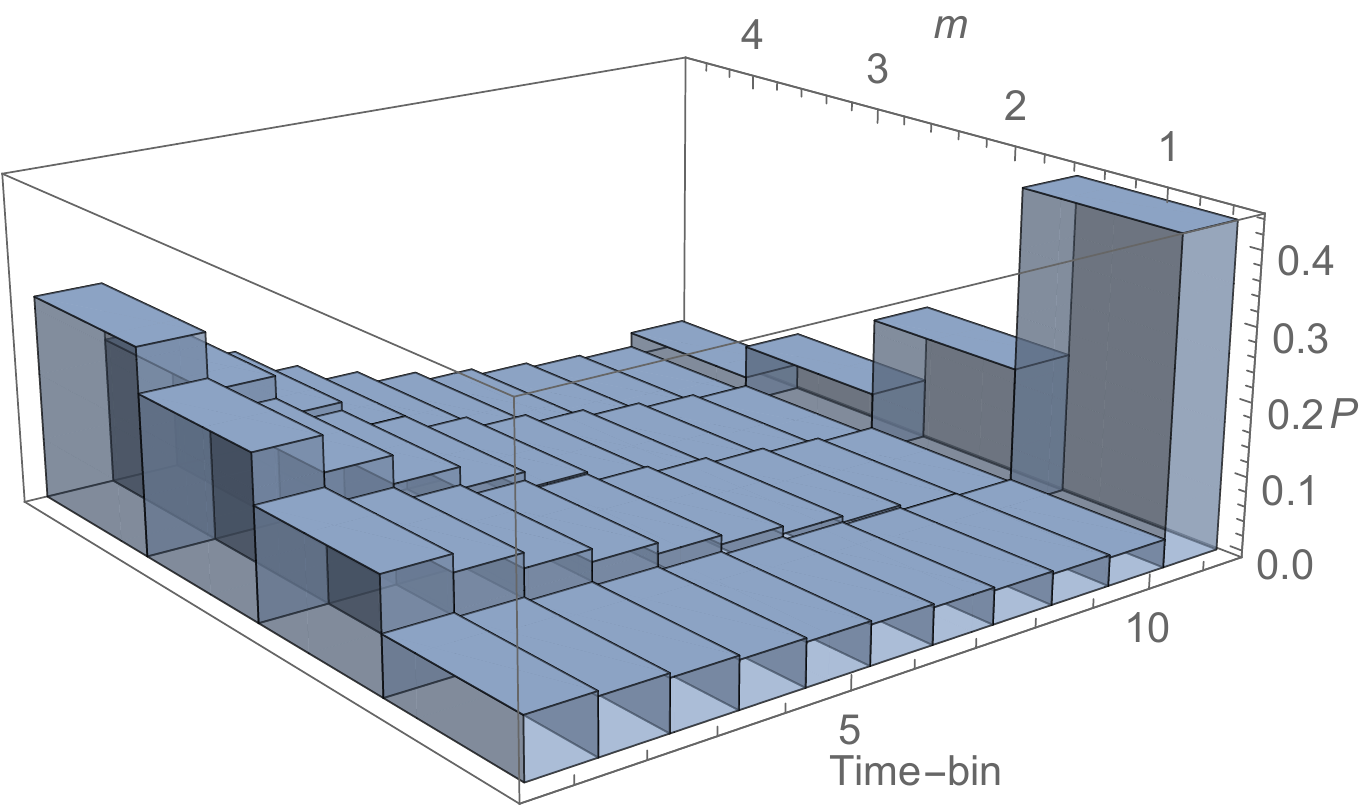} 
\caption{The probability distribution of last heralded photons in 10 time-bins for using $m$ sources in parallel. The time-bins are enumerated by the number of passes through the loop. The probability indicated in position 11 is the probability of no heralding occurring. As more sources are used in parallel the probability of the no heralding event is reduced and the probability that a photon in a low-numbered time-bin is emitted is increased.  In this example $\bar{n}=0.1$ and $\eta=0.95$.  } \label{fig:probsmultiple}
\end{figure}

The resultant joint probability distribution for $m$ sources running in parallel, where we take the source with the last heralded photon across all $m$ sources as the output, can be written in a form suitable for numerical computation:
\begin{equation}
p_m(u) = \sum_{j_1\ldots j_m=0}^t T(u,j_1 \ldots j_m) \prod_{j\in\{j_1 \ldots j_m\}} p(j),
\end{equation}
where
\begin{equation}\label{eq:probaiities3}
T(u,j_1\ldots j_m) = \begin{cases}
1  & \min(j_1,j_2,\ldots j_m)=u\\
0 & \text{otherwise}
\end{cases}.
\end{equation}
The distributions for using up to four parallel sources are shown in Fig.~\ref{fig:probsmultiple}.

\begin{figure}[htb]
\centering
\includegraphics[width=0.9\columnwidth]{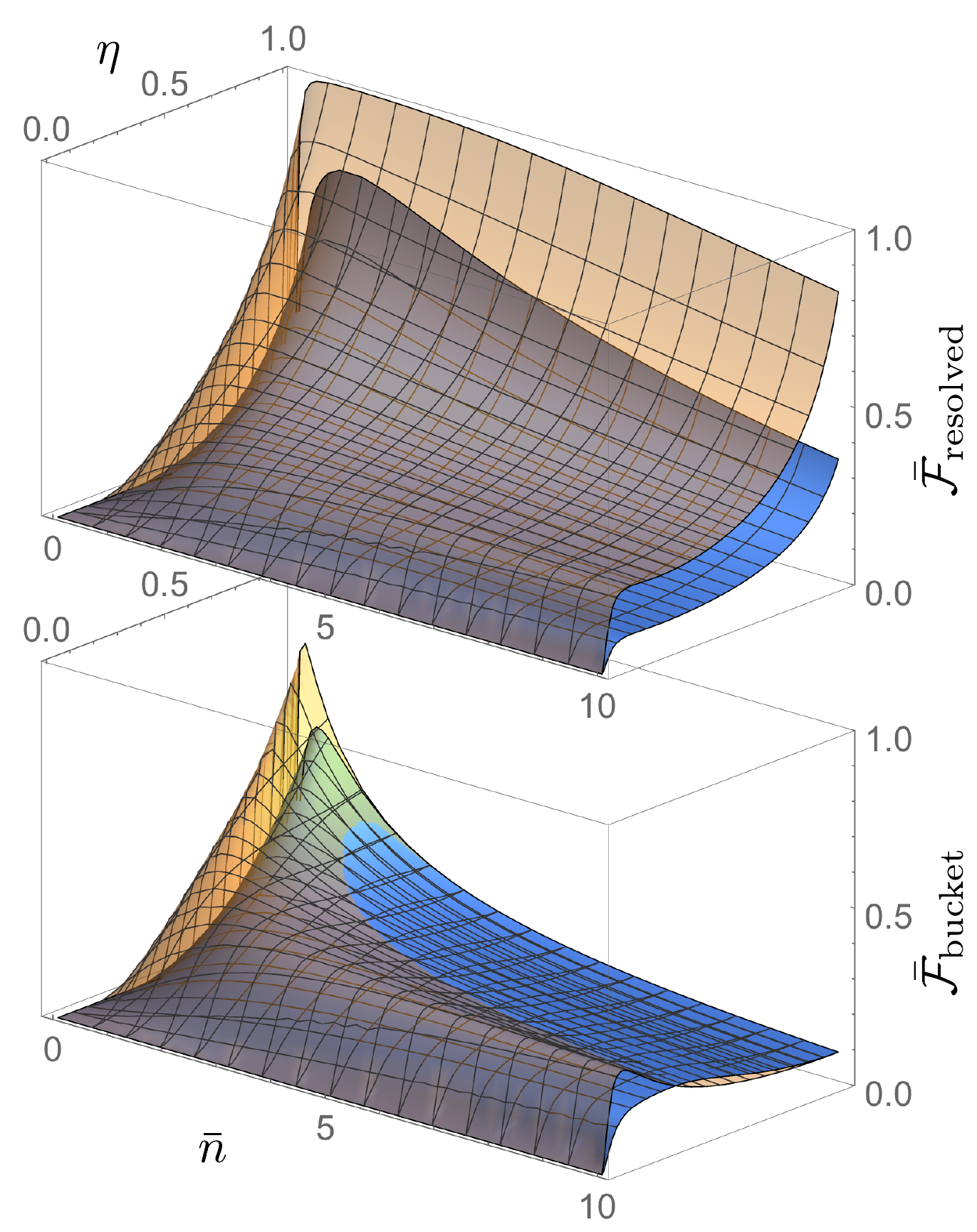}
\caption{Improvement obtained by running four loop-sources in parallel and selecting the photons from the latest time-bins in each output pulse. Solid blue surfaces are for a single loop source, translucent yellow surfaces are for four parallel sources. Top graph is using resolved detectors while bottom graph is bucket detectors. For all configurations only five time-bins where used ($t=5$) in each source.} \label{fig:4sourceF}
\end{figure}

Using these probability distributions it is straightforward to calculate the average unconditional fidelities. As an illustrative example, consider a loop-source that only has five time-bins. The improvement possible by using up to four such sources in parallel is depicted in Fig.~\ref{fig:4sourceF}. 

Of course these are the maximum fidelities we could expect, as we have not taken into account the losses inherent in any spatial switching and routing scheme. various strategies for spatial multiplexing have been previously studied \cite{bib:MosleyMultiplex,bib:ObrienMultiplex, 13mazzarella023848,bib:JenneweinBarbieriWhite}. Such strategies may be readily adapted to improving the quality of a bank of parallel loop sources. 

It is interesting to note that in the bucket detector there are parameter regimes where combining multiple parallel sources fares worse than just taking the output of a single source. Combining sources by picking the latest heralded time-bin reduces the effective average loss. The parameter regimes where this strategy is a disadvantage are the regimes where loss is helping improve the fidelity. The bucket heralding detector cannot distinguish between numbers of photons and so lets through higher order photon states, and with sufficient loss some of these states are reduced to single photon states on output.

The above observation suggests that it might be beneficial to admit a certain fraction of two-photon states when using the resolved detectors. These states would help mitigate the effects of loss in the switching and memory. The fraction of these higher-order terms accepted can be optimised over where in the pulse train they occur and we envision quite sophisticated switching and parallel-source combination strategies are possible to dynamically increase the quality of the output. Although outside the scope of the this document and reserved for future work, all of this is possible because the loop source produces a quality parameter for each output in the form of the heralding signature. 

\section{Experimental viability} \label{sec:ex_params}

Finally, let us consider the feasibility of our design with current and near-term technology, in terms of key experimental parameters including switch and fibre loss, pulse spreading and switching rates. We concentrate on the telecommunication band at 1550nm, where fibre-based technology is most favourable.

In general, single-photon multiplexing is only worthwhile if the individual sources produce heralded single photons of near perfect purity so that only one spectral mode is present. (This situation corresponds to the thermal probability distribution used earlier in the paper). In current 1550~nm silicon-based photon pair sources, this typically translates to photons with coherence time of order \mbox{$10-20\mathrm{ ps}$} and a spatial extent of a few mm \cite{Takesue2007, bib:CollinsMultiplex, Silverstone2014}. 

Detector efficiency and counting rate are also critical and there is typically a trade-off in these parameters. In the 1550~nm band, commercial InGaAs APD-based detectors operate with around \mbox{$\eta_d=0.3$} efficiency and detection rates of 100~MHz or better. Superconducting nanowire detectors offer much lower dark-counts avoiding an inflated heralding probability. Recent devices are also showing much enhanced efficiency with values of \mbox{$\eta_d \gtrsim 0.8$} \cite{marsili2013}, and while the maximum detection rates were previously in the MHz range, they can now exceed 10MHz \cite{Yamashita:13}.

\begin{figure}[htb]
\includegraphics[width=\columnwidth]{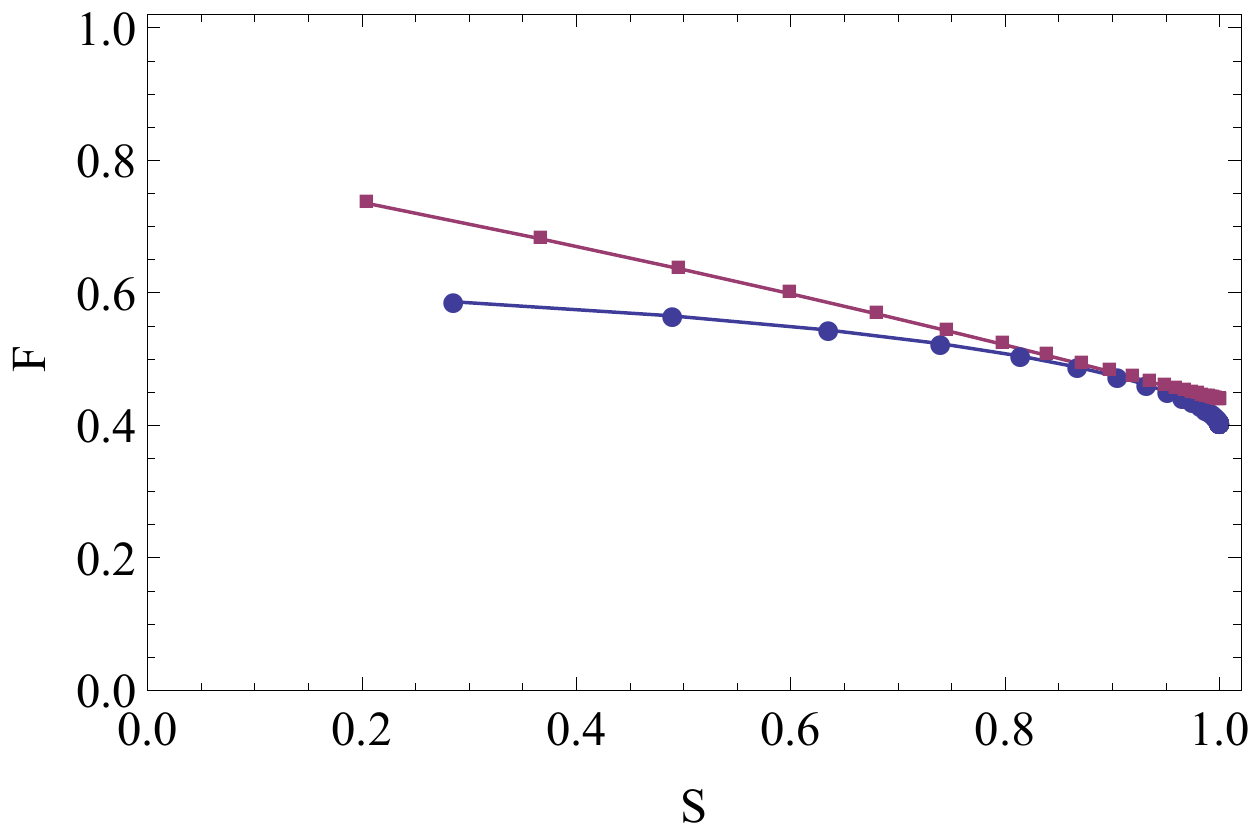}
\caption{Fidelity versus heralding success probability tradeoff, parameterised against the number of time-bins (increasing from left to right). Here \mbox{$\eta_f=1, \eta_d=0.8, \eta_s=0.8, \bar{n}=0.5$}, reflective of the parameters discussed in Sec. \ref{sec:ex_params}. (blue) bucket detectors, (red) number-resolved detectors.} \label{fig:FS_exp}
\end{figure}

To avoid pulse overlap, one might then decide (fairly conservatively) that successive photons should be spaced no closer than 100~ps. This corresponds to a pump pulse repetition rate of 10~GHz, somewhat higher than today's pair sources, and a fibre-loop length of only 2~cm, which would be impractically short. Instead, we consider a pump pulse repetition rate 1~GHz, consistent with several very recent sources \cite{Morris2014, Jin2014}. Assuming mean photon rates of $\bar{n}\lesssim 0.1$, this is also consistent with state-of-the-art ceramic waveguide switches which can operate at a few MHz and have already been used in spatial photon multiplexing experiments \cite{bib:CollinsMultiplex}. Such switches currently exhibit transmission of order \mbox{$\eta_s\approx 0.8$}. Significant additional improvements in transmission will probably be gradual at best. In this regime fibre losses are minimal, thus \mbox{$\eta_f\approx 1$}. For \mbox{$t=10$} orbits around the loop, the net transmission is \mbox{$\tau_t\approx\eta_s^{t+1}\approx 0.1$}, at which point there is marginal benefit in utilising more round-trips. With the required loop delay of 10~ns, the fibre length is only 2.0~m and loop losses are negligible (below 0.01\% for 10 orbits) \cite{AgrawalNFO}. Similarly the dispersion length for significant pulse spreading is of order 20~km, and so irrelevant. For the foreseeable future then, the loop performance is entirely dominated by the switch transmission. A figure representative of these parameters is shown in Fig.~\ref{fig:FS_exp}, showing the tradeoff between fidelity and heralding success probability, parameterised by the number of time-bins.

Another common wavelength band of interest for pair generation is around 800~nm where silicon avalanche photodiode detectors (APDs) offer much greater efficiency of \mbox{$\eta_d>0.7$}. For most parameters discussed above, the conclusions are largely unchanged. The fibre attenuation is just 3~dB/km, for which fibre losses are negligible. However, the technology for high speed waveguide switches in this wavelength band is less mature and employing fewer round-trips of the loop is favourable.

As a more generally available scenario using more standard pump sources, many mode-locked lasers run at an 80~MHz repetition rate, where successive pulses appear at 12.5~ns. This is a rate more compatible with the counting rates supported by contemporary detectors as we discuss in a moment. This pulse spacing corresponds to a fibre length of 2.6~m. Even then, the attenuation of \mbox{$t=10$} loops is of order 0.1\% at 1550~nm, and only 3\% at 800~nm, with pulse spreading still irrelevant.

Number-resolving detectors based on transition edge sensors (TES) have also now been demonstrated with efficiencies of \mbox{$\eta_d>0.8$} at 1550~nm, but the count rates are limited to around 100~kHz \cite{Calkins:13}. Using such detectors would then necessitate a matching drop in the pump pulse rate to 100~kHz. The corresponding fibre loop length of 2.1~km then gives a transmission of \mbox{$\eta_f=0.91$} and the fibre and switch efficiencies begin to be comparable.

\section{Conclusion}

We have analysed a scheme for multiplexing SPDC sources using temporal encoding, to allow the construction of a source with higher heralding efficiency. The scheme requires a single SPDC source, a dynamic switch, a fibre-loop, and a time-resolved photo-detector. Unlike spatial multiplexing schemes, the complexity of the scheme presented here does not increase with the level of multiplexing, making it very favourable in terms of experimental resource requirements. The heralding efficiency of the system may be made asymptotically close to unity when enough time-bins are used. However, doing so comes at the expense of fidelity, which degrades as the number of time-bins increases. The technologies required for elementary demonstrations of this protocol are largely available today, and could substantially improve the heralding efficiency of an SPDC-based source using present-day technology.

\begin{acknowledgments}
We thank Keith Motes, Jonathan Dowling \& Michael Bremner for helpful discussions. This research was conducted by the Australian Research Council Centre of Excellence for Engineered Quantum Systems (Project number CE110001013). PPR acknowledges financial support from Lockheed Martin.
\end{acknowledgments}

\bibliography{bibliography,mikebib}

\begin{thebibliography}{43}
\expandafter\ifx\csname natexlab\endcsname\relax\def\natexlab#1{#1}\fi
\expandafter\ifx\csname bibnamefont\endcsname\relax
  \def\bibnamefont#1{#1}\fi
\expandafter\ifx\csname bibfnamefont\endcsname\relax
  \def\bibfnamefont#1{#1}\fi
\expandafter\ifx\csname citenamefont\endcsname\relax
  \def\citenamefont#1{#1}\fi
\expandafter\ifx\csname url\endcsname\relax
  \def\url#1{\texttt{#1}}\fi
\expandafter\ifx\csname urlprefix\endcsname\relax\def\urlprefix{URL }\fi
\providecommand{\bibinfo}[2]{#2}
\providecommand{\eprint}[2][]{\url{#2}}

\bibitem[{\citenamefont{Knill et~al.}(2001)\citenamefont{Knill, Laflamme, and
  Milburn}}]{bib:KLM01}
\bibinfo{author}{\bibfnamefont{E.}~\bibnamefont{Knill}},
  \bibinfo{author}{\bibfnamefont{R.}~\bibnamefont{Laflamme}}, \bibnamefont{and}
  \bibinfo{author}{\bibfnamefont{G.}~\bibnamefont{Milburn}},
  \bibinfo{journal}{Nature (London)} \textbf{\bibinfo{volume}{409}},
  \bibinfo{pages}{46} (\bibinfo{year}{2001}).

\bibitem[{\citenamefont{Aaronson and Arkhipov}(2011)}]{bib:AaronsonArkhipov10}
\bibinfo{author}{\bibfnamefont{S.}~\bibnamefont{Aaronson}} \bibnamefont{and}
  \bibinfo{author}{\bibfnamefont{A.}~\bibnamefont{Arkhipov}},
  \bibinfo{journal}{Proc. ACM STOC (New York)} p. \bibinfo{pages}{333}
  (\bibinfo{year}{2011}).

\bibitem[{\citenamefont{Motes et~al.}(2015)\citenamefont{Motes, Olson, Rabeaux,
  Dowling, Olson, and Rohde}}]{bib:MORDOR}
\bibinfo{author}{\bibfnamefont{K.~R.} \bibnamefont{Motes}},
  \bibinfo{author}{\bibfnamefont{J.~P.} \bibnamefont{Olson}},
  \bibinfo{author}{\bibfnamefont{E.~J.} \bibnamefont{Rabeaux}},
  \bibinfo{author}{\bibfnamefont{J.~P.} \bibnamefont{Dowling}},
  \bibinfo{author}{\bibfnamefont{S.~J.} \bibnamefont{Olson}}, \bibnamefont{and}
  \bibinfo{author}{\bibfnamefont{P.~P.} \bibnamefont{Rohde}}
  (\bibinfo{year}{2015}), \eprint{arXiv:1501.01067}.

\bibitem[{\citenamefont{Adam et~al.}(2014)\citenamefont{Adam, Mechler, Santa,
  and Koniorczyk}}]{14adam053834}
\bibinfo{author}{\bibfnamefont{P.}~\bibnamefont{Adam}},
  \bibinfo{author}{\bibfnamefont{M.}~\bibnamefont{Mechler}},
  \bibinfo{author}{\bibfnamefont{I.}~\bibnamefont{Santa}}, \bibnamefont{and}
  \bibinfo{author}{\bibfnamefont{M.}~\bibnamefont{Koniorczyk}},
  \bibinfo{journal}{Phys. Rev. A} \textbf{\bibinfo{volume}{90}},
  \bibinfo{pages}{053834} (\bibinfo{year}{2014}).

\bibitem[{\citenamefont{Francis-Jones and Mosley}(2014)}]{bib:MosleyMultiplex}
\bibinfo{author}{\bibfnamefont{R.~J.~A.} \bibnamefont{Francis-Jones}}
  \bibnamefont{and} \bibinfo{author}{\bibfnamefont{P.~J.} \bibnamefont{Mosley}}
  (\bibinfo{year}{2014}), \eprint{arXiv:1409.1394}.

\bibitem[{\citenamefont{Bonneau et~al.}(2014)\citenamefont{Bonneau, Mendoza,
  O'Brien, and Thompson}}]{bib:ObrienMultiplex}
\bibinfo{author}{\bibfnamefont{D.}~\bibnamefont{Bonneau}},
  \bibinfo{author}{\bibfnamefont{G.~J.} \bibnamefont{Mendoza}},
  \bibinfo{author}{\bibfnamefont{J.~L.} \bibnamefont{O'Brien}},
  \bibnamefont{and} \bibinfo{author}{\bibfnamefont{M.~G.}
  \bibnamefont{Thompson}} (\bibinfo{year}{2014}), \eprint{arXiv:1409.5341}.

\bibitem[{\citenamefont{Meany et~al.}(2014)\citenamefont{Meany, Ngah, Collins,
  Clark, Williams, Eggleton, Steel, Withford, Alibart, and
  Tanzilli}}]{14meany42}
\bibinfo{author}{\bibfnamefont{T.}~\bibnamefont{Meany}},
  \bibinfo{author}{\bibfnamefont{L.~A.} \bibnamefont{Ngah}},
  \bibinfo{author}{\bibfnamefont{M.~J.} \bibnamefont{Collins}},
  \bibinfo{author}{\bibfnamefont{A.~S.} \bibnamefont{Clark}},
  \bibinfo{author}{\bibfnamefont{R.~J.} \bibnamefont{Williams}},
  \bibinfo{author}{\bibfnamefont{B.~J.} \bibnamefont{Eggleton}},
  \bibinfo{author}{\bibfnamefont{M.~J.} \bibnamefont{Steel}},
  \bibinfo{author}{\bibfnamefont{M.~J.} \bibnamefont{Withford}},
  \bibinfo{author}{\bibfnamefont{O.}~\bibnamefont{Alibart}}, \bibnamefont{and}
  \bibinfo{author}{\bibfnamefont{S.}~\bibnamefont{Tanzilli}},
  \bibinfo{journal}{Laser {\&} Photonics Reviews} \textbf{\bibinfo{volume}{8}},
  \bibinfo{pages}{L42} (\bibinfo{year}{2014}).

\bibitem[{\citenamefont{Collins et~al.}(2013)\citenamefont{Collins, Xiong, Rey,
  Vo, He, Shahnia, Reardon, Krauss, Steel, Clark
  et~al.}}]{bib:CollinsMultiplex}
\bibinfo{author}{\bibfnamefont{M.~J.} \bibnamefont{Collins}},
  \bibinfo{author}{\bibfnamefont{C.}~\bibnamefont{Xiong}},
  \bibinfo{author}{\bibfnamefont{I.~H.} \bibnamefont{Rey}},
  \bibinfo{author}{\bibfnamefont{T.~D.} \bibnamefont{Vo}},
  \bibinfo{author}{\bibfnamefont{J.}~\bibnamefont{He}},
  \bibinfo{author}{\bibfnamefont{S.}~\bibnamefont{Shahnia}},
  \bibinfo{author}{\bibfnamefont{C.}~\bibnamefont{Reardon}},
  \bibinfo{author}{\bibfnamefont{T.~F.} \bibnamefont{Krauss}},
  \bibinfo{author}{\bibfnamefont{M.~J.} \bibnamefont{Steel}},
  \bibinfo{author}{\bibfnamefont{A.~S.} \bibnamefont{Clark}},
  \bibnamefont{et~al.}, \bibinfo{journal}{Nature Comm.}
  \textbf{\bibinfo{volume}{4}}, \bibinfo{pages}{2852} (\bibinfo{year}{2013}).

\bibitem[{\citenamefont{Mazzarella et~al.}(2013)\citenamefont{Mazzarella,
  Ticozzi, Sergienko, Vallone, and Villoresi}}]{13mazzarella023848}
\bibinfo{author}{\bibfnamefont{L.}~\bibnamefont{Mazzarella}},
  \bibinfo{author}{\bibfnamefont{F.}~\bibnamefont{Ticozzi}},
  \bibinfo{author}{\bibfnamefont{A.~V.} \bibnamefont{Sergienko}},
  \bibinfo{author}{\bibfnamefont{G.}~\bibnamefont{Vallone}}, \bibnamefont{and}
  \bibinfo{author}{\bibfnamefont{P.}~\bibnamefont{Villoresi}},
  \bibinfo{journal}{Physical Review A} \textbf{\bibinfo{volume}{88}},
  \bibinfo{pages}{023848} (\bibinfo{year}{2013}).

\bibitem[{\citenamefont{Christ and Silberhorn}(2012)}]{bib:ChristMultiplex}
\bibinfo{author}{\bibfnamefont{A.}~\bibnamefont{Christ}} \bibnamefont{and}
  \bibinfo{author}{\bibfnamefont{C.}~\bibnamefont{Silberhorn}},
  \bibinfo{journal}{Phys. Rev. A} \textbf{\bibinfo{volume}{85}},
  \bibinfo{pages}{023829} (\bibinfo{year}{2012}).

\bibitem[{\citenamefont{Ma et~al.}(2011)\citenamefont{Ma, Zotter, Kofler,
  Jennewein, and Zeilinger}}]{bib:ZeilingerMultiplex}
\bibinfo{author}{\bibfnamefont{X.-s.} \bibnamefont{Ma}},
  \bibinfo{author}{\bibfnamefont{S.}~\bibnamefont{Zotter}},
  \bibinfo{author}{\bibfnamefont{J.}~\bibnamefont{Kofler}},
  \bibinfo{author}{\bibfnamefont{T.}~\bibnamefont{Jennewein}},
  \bibnamefont{and}
  \bibinfo{author}{\bibfnamefont{A.}~\bibnamefont{Zeilinger}},
  \bibinfo{journal}{Phys. Rev. A} \textbf{\bibinfo{volume}{83}},
  \bibinfo{pages}{043814} (\bibinfo{year}{2011}).

\bibitem[{\citenamefont{Jennewein et~al.}(2011)\citenamefont{Jennewein,
  Barbieri, and White}}]{bib:JenneweinBarbieriWhite}
\bibinfo{author}{\bibfnamefont{T.}~\bibnamefont{Jennewein}},
  \bibinfo{author}{\bibfnamefont{M.}~\bibnamefont{Barbieri}}, \bibnamefont{and}
  \bibinfo{author}{\bibfnamefont{A.~G.} \bibnamefont{White}},
  \bibinfo{journal}{J. Mod. Opt.} \textbf{\bibinfo{volume}{58}},
  \bibinfo{pages}{276} (\bibinfo{year}{2011}).

\bibitem[{\citenamefont{Shapiro and Wong}(2007)}]{bib:ShapiroMultiplex}
\bibinfo{author}{\bibfnamefont{J.~H.} \bibnamefont{Shapiro}} \bibnamefont{and}
  \bibinfo{author}{\bibfnamefont{F.~N.} \bibnamefont{Wong}},
  \bibinfo{journal}{Opt. Lett.} \textbf{\bibinfo{volume}{32}},
  \bibinfo{pages}{2698} (\bibinfo{year}{2007}).

\bibitem[{\citenamefont{Migdall et~al.}(2002)\citenamefont{Migdall, Branning,
  and Castelletto}}]{bib:MigdallMultiplex}
\bibinfo{author}{\bibfnamefont{A.}~\bibnamefont{Migdall}},
  \bibinfo{author}{\bibfnamefont{D.}~\bibnamefont{Branning}}, \bibnamefont{and}
  \bibinfo{author}{\bibfnamefont{S.}~\bibnamefont{Castelletto}},
  \bibinfo{journal}{Phys. Rev. A} \textbf{\bibinfo{volume}{66}},
  \bibinfo{pages}{053805} (\bibinfo{year}{2002}).

\bibitem[{\citenamefont{Mendoza et~al.}(2015)\citenamefont{Mendoza, Santagati,
  Munns, Hemsley, Piekarek, Martin-Lopez, Marshall, Bonneau, Thompson, and
  O'Brien}}]{15mendoza}
\bibinfo{author}{\bibfnamefont{G.~J.} \bibnamefont{Mendoza}},
  \bibinfo{author}{\bibfnamefont{R.}~\bibnamefont{Santagati}},
  \bibinfo{author}{\bibfnamefont{J.}~\bibnamefont{Munns}},
  \bibinfo{author}{\bibfnamefont{E.}~\bibnamefont{Hemsley}},
  \bibinfo{author}{\bibfnamefont{M.}~\bibnamefont{Piekarek}},
  \bibinfo{author}{\bibfnamefont{E.}~\bibnamefont{Martin-Lopez}},
  \bibinfo{author}{\bibfnamefont{G.~D.} \bibnamefont{Marshall}},
  \bibinfo{author}{\bibfnamefont{D.}~\bibnamefont{Bonneau}},
  \bibinfo{author}{\bibfnamefont{M.~G.} \bibnamefont{Thompson}},
  \bibnamefont{and} \bibinfo{author}{\bibfnamefont{J.~L.}
  \bibnamefont{O'Brien}} (\bibinfo{year}{2015}), \eprint{1503.01215}.

\bibitem[{\citenamefont{Schmiegelow and Larotonda}(2013)}]{13schmiegelow447}
\bibinfo{author}{\bibfnamefont{C.~T.} \bibnamefont{Schmiegelow}}
  \bibnamefont{and} \bibinfo{author}{\bibfnamefont{M.~A.}
  \bibnamefont{Larotonda}}, \bibinfo{journal}{Appl. Phys. B}
  \textbf{\bibinfo{volume}{116}}, \bibinfo{pages}{447} (\bibinfo{year}{2013}).

\bibitem[{\citenamefont{Glebov et~al.}(2013)\citenamefont{Glebov, Fan, and
  Migdall}}]{13glebov031115}
\bibinfo{author}{\bibfnamefont{B.~L.} \bibnamefont{Glebov}},
  \bibinfo{author}{\bibfnamefont{J.}~\bibnamefont{Fan}}, \bibnamefont{and}
  \bibinfo{author}{\bibfnamefont{A.}~\bibnamefont{Migdall}},
  \bibinfo{journal}{Appl. Phys. Lett.} \textbf{\bibinfo{volume}{103}},
  \bibinfo{pages}{031115} (\bibinfo{year}{2013}).

\bibitem[{\citenamefont{Mower and Englund}(2011)}]{11mower052326}
\bibinfo{author}{\bibfnamefont{J.}~\bibnamefont{Mower}} \bibnamefont{and}
  \bibinfo{author}{\bibfnamefont{D.}~\bibnamefont{Englund}},
  \bibinfo{journal}{Phys. Rev. A} \textbf{\bibinfo{volume}{84}},
  \bibinfo{pages}{052326} (\bibinfo{year}{2011}).

\bibitem[{\citenamefont{McCusker et~al.}(2008)\citenamefont{McCusker, Peters,
  VanDevender, and Kwiat}}]{08mccuskerjtua117}
\bibinfo{author}{\bibfnamefont{K.~T.} \bibnamefont{McCusker}},
  \bibinfo{author}{\bibfnamefont{N.~A.} \bibnamefont{Peters}},
  \bibinfo{author}{\bibfnamefont{A.~P.} \bibnamefont{VanDevender}},
  \bibnamefont{and} \bibinfo{author}{\bibfnamefont{P.~G.} \bibnamefont{Kwiat}},
  in \emph{\bibinfo{booktitle}{2008 Conference on Lasers and Electro-Optics}}
  (\bibinfo{publisher}{IEEE}, \bibinfo{year}{2008}), pp. \bibinfo{pages}{1--2}.

\bibitem[{\citenamefont{Peters et~al.}(2006)\citenamefont{Peters, Arnold,
  VanDevender, Jeffrey, Rangarajan, Hosten, Barreiro, Altepeter, and
  KWIAT}}]{06peters630507}
\bibinfo{author}{\bibfnamefont{N.~A.} \bibnamefont{Peters}},
  \bibinfo{author}{\bibfnamefont{K.~J.} \bibnamefont{Arnold}},
  \bibinfo{author}{\bibfnamefont{A.~P.} \bibnamefont{VanDevender}},
  \bibinfo{author}{\bibfnamefont{E.~R.} \bibnamefont{Jeffrey}},
  \bibinfo{author}{\bibfnamefont{R.}~\bibnamefont{Rangarajan}},
  \bibinfo{author}{\bibfnamefont{O.}~\bibnamefont{Hosten}},
  \bibinfo{author}{\bibfnamefont{J.~T.} \bibnamefont{Barreiro}},
  \bibinfo{author}{\bibfnamefont{J.~B.} \bibnamefont{Altepeter}},
  \bibnamefont{and} \bibinfo{author}{\bibfnamefont{P.~G.} \bibnamefont{KWIAT}},
  in \emph{\bibinfo{booktitle}{Proceedings of SPIE}} (\bibinfo{year}{2006}), p.
  \bibinfo{pages}{630507}.

\bibitem[{\citenamefont{Jeffrey et~al.}(2004)\citenamefont{Jeffrey, Peters, and
  Kwiat}}]{04jeffrey1}
\bibinfo{author}{\bibfnamefont{E.}~\bibnamefont{Jeffrey}},
  \bibinfo{author}{\bibfnamefont{N.~A.} \bibnamefont{Peters}},
  \bibnamefont{and} \bibinfo{author}{\bibfnamefont{P.~G.} \bibnamefont{Kwiat}},
  \bibinfo{journal}{New J. of Phy.} \textbf{\bibinfo{volume}{6}},
  \bibinfo{pages}{1} (\bibinfo{year}{2004}).

\bibitem[{\citenamefont{Pittman et~al.}(2004)\citenamefont{Pittman, Fitch,
  Jacobs, and Franson}}]{04pittman57}
\bibinfo{author}{\bibfnamefont{T.~B.} \bibnamefont{Pittman}},
  \bibinfo{author}{\bibfnamefont{M.~J.} \bibnamefont{Fitch}},
  \bibinfo{author}{\bibfnamefont{B.~C.} \bibnamefont{Jacobs}},
  \bibnamefont{and} \bibinfo{author}{\bibfnamefont{J.~D.}
  \bibnamefont{Franson}}, in \emph{\bibinfo{booktitle}{Optical Science and
  Technology, SPIE's 48th Annual Meeting}}, edited by
  \bibinfo{editor}{\bibfnamefont{R.~E.} \bibnamefont{Meyers}} \bibnamefont{and}
  \bibinfo{editor}{\bibfnamefont{Y.}~\bibnamefont{Shih}}
  (\bibinfo{publisher}{SPIE}, \bibinfo{year}{2004}), pp.
  \bibinfo{pages}{57--65}.

\bibitem[{\citenamefont{Pittman et~al.}(2002)\citenamefont{Pittman, Jacobs, and
  Franson}}]{bib:PittmanSPDC}
\bibinfo{author}{\bibfnamefont{T.~B.} \bibnamefont{Pittman}},
  \bibinfo{author}{\bibfnamefont{B.~C.} \bibnamefont{Jacobs}},
  \bibnamefont{and} \bibinfo{author}{\bibfnamefont{J.~D.}
  \bibnamefont{Franson}}, \bibinfo{journal}{Phys. Rev. A}
  \textbf{\bibinfo{volume}{66}}, \bibinfo{pages}{042303}
  (\bibinfo{year}{2002}).

\bibitem[{\citenamefont{Nunn et~al.}(2013)\citenamefont{Nunn, Langford,
  Kolthammer, Champion, Sprague, Michelberger, Jin, England, and
  Walmsley}}]{bib:Nunn13}
\bibinfo{author}{\bibfnamefont{J.}~\bibnamefont{Nunn}},
  \bibinfo{author}{\bibfnamefont{N.~K.} \bibnamefont{Langford}},
  \bibinfo{author}{\bibfnamefont{W.~S.} \bibnamefont{Kolthammer}},
  \bibinfo{author}{\bibfnamefont{T.~F.~M.} \bibnamefont{Champion}},
  \bibinfo{author}{\bibfnamefont{M.~R.} \bibnamefont{Sprague}},
  \bibinfo{author}{\bibfnamefont{P.~S.} \bibnamefont{Michelberger}},
  \bibinfo{author}{\bibfnamefont{X.-M.} \bibnamefont{Jin}},
  \bibinfo{author}{\bibfnamefont{D.~G.} \bibnamefont{England}},
  \bibnamefont{and} \bibinfo{author}{\bibfnamefont{I.~A.}
  \bibnamefont{Walmsley}}, \bibinfo{journal}{Phys. Rev. Lett.}
  \textbf{\bibinfo{volume}{110}}, \bibinfo{pages}{133601}
  (\bibinfo{year}{2013}).

\bibitem[{\citenamefont{Motes et~al.}(2014)\citenamefont{Motes, Gilchrist,
  Dowling, and Rohde}}]{bib:MotesLoop}
\bibinfo{author}{\bibfnamefont{K.~R.} \bibnamefont{Motes}},
  \bibinfo{author}{\bibfnamefont{A.}~\bibnamefont{Gilchrist}},
  \bibinfo{author}{\bibfnamefont{J.~P.} \bibnamefont{Dowling}},
  \bibnamefont{and} \bibinfo{author}{\bibfnamefont{P.~P.} \bibnamefont{Rohde}},
  \bibinfo{journal}{Phys. Rev. Lett.} \textbf{\bibinfo{volume}{113}},
  \bibinfo{pages}{120501} (\bibinfo{year}{2014}).

\bibitem[{\citenamefont{Rohde}(2015)}]{bib:RohdeLoopLOQC}
\bibinfo{author}{\bibfnamefont{P.~P.} \bibnamefont{Rohde}},
  \bibinfo{journal}{Phys. Rev. A} \textbf{\bibinfo{volume}{91}},
  \bibinfo{pages}{012306} (\bibinfo{year}{2015}).

\bibitem[{\citenamefont{Fitch et~al.}(2003)\citenamefont{Fitch, Jacobs,
  Pittman, and Franson}}]{bib:Fitch03}
\bibinfo{author}{\bibfnamefont{M.~J.} \bibnamefont{Fitch}},
  \bibinfo{author}{\bibfnamefont{B.~C.} \bibnamefont{Jacobs}},
  \bibinfo{author}{\bibfnamefont{T.~B.} \bibnamefont{Pittman}},
  \bibnamefont{and} \bibinfo{author}{\bibfnamefont{J.~D.}
  \bibnamefont{Franson}}, \bibinfo{journal}{Phys. Rev. A}
  \textbf{\bibinfo{volume}{68}}, \bibinfo{pages}{043814}
  (\bibinfo{year}{2003}).

\bibitem[{\citenamefont{Achilles et~al.}(2004)\citenamefont{Achilles,
  Silberhorn, Sliwa, Banaszek, Walmsley, Fitch, Jacobs, Pittman, and
  Franson}}]{bib:Achilles04}
\bibinfo{author}{\bibfnamefont{D.}~\bibnamefont{Achilles}},
  \bibinfo{author}{\bibfnamefont{C.}~\bibnamefont{Silberhorn}},
  \bibinfo{author}{\bibfnamefont{C.}~\bibnamefont{Sliwa}},
  \bibinfo{author}{\bibfnamefont{K.}~\bibnamefont{Banaszek}},
  \bibinfo{author}{\bibfnamefont{I.~A.} \bibnamefont{Walmsley}},
  \bibinfo{author}{\bibfnamefont{M.~J.} \bibnamefont{Fitch}},
  \bibinfo{author}{\bibfnamefont{B.~C.} \bibnamefont{Jacobs}},
  \bibinfo{author}{\bibfnamefont{T.~B.} \bibnamefont{Pittman}},
  \bibnamefont{and} \bibinfo{author}{\bibfnamefont{J.~D.}
  \bibnamefont{Franson}}, \bibinfo{journal}{J. Mod. Opt.}
  \textbf{\bibinfo{volume}{51}}, \bibinfo{pages}{1499} (\bibinfo{year}{2004}).

\bibitem[{\citenamefont{Rohde et~al.}(2007)\citenamefont{Rohde, Webb,
  Huntington, and Ralph}}]{bib:RohdeWebb07}
\bibinfo{author}{\bibfnamefont{P.~P.} \bibnamefont{Rohde}},
  \bibinfo{author}{\bibfnamefont{J.~G.} \bibnamefont{Webb}},
  \bibinfo{author}{\bibfnamefont{E.~H.} \bibnamefont{Huntington}},
  \bibnamefont{and} \bibinfo{author}{\bibfnamefont{T.~C.} \bibnamefont{Ralph}},
  \bibinfo{journal}{New J. Phys.} \textbf{\bibinfo{volume}{9}},
  \bibinfo{pages}{233} (\bibinfo{year}{2007}).

\bibitem[{\citenamefont{Schreiber et~al.}(2010)\citenamefont{Schreiber,
  Cassemiro, Poto{\u c}ek, G{\' a}bris, Mosley, Andersson, Jex, and
  Silberhorn}}]{bib:Schreiber10}
\bibinfo{author}{\bibfnamefont{A.}~\bibnamefont{Schreiber}},
  \bibinfo{author}{\bibfnamefont{K.~N.} \bibnamefont{Cassemiro}},
  \bibinfo{author}{\bibfnamefont{V.}~\bibnamefont{Poto{\u c}ek}},
  \bibinfo{author}{\bibfnamefont{A.}~\bibnamefont{G{\' a}bris}},
  \bibinfo{author}{\bibfnamefont{P.~J.} \bibnamefont{Mosley}},
  \bibinfo{author}{\bibfnamefont{E.}~\bibnamefont{Andersson}},
  \bibinfo{author}{\bibfnamefont{I.}~\bibnamefont{Jex}}, \bibnamefont{and}
  \bibinfo{author}{\bibfnamefont{C.}~\bibnamefont{Silberhorn}},
  \bibinfo{journal}{Phys. Rev. Lett.} \textbf{\bibinfo{volume}{104}},
  \bibinfo{pages}{050502} (\bibinfo{year}{2010}).

\bibitem[{\citenamefont{Schreiber et~al.}(2012)\citenamefont{Schreiber,
  G{\'a}bris, Rohde, Laiho, {\v S}tefa{\v n}{\' a}k, Poto{\u c}ek, Jex, and
  Silberhorn}}]{bib:Schreiber12}
\bibinfo{author}{\bibfnamefont{A.}~\bibnamefont{Schreiber}},
  \bibinfo{author}{\bibfnamefont{A.}~\bibnamefont{G{\'a}bris}},
  \bibinfo{author}{\bibfnamefont{P.~P.} \bibnamefont{Rohde}},
  \bibinfo{author}{\bibfnamefont{K.}~\bibnamefont{Laiho}},
  \bibinfo{author}{\bibfnamefont{M.}~\bibnamefont{{\v S}tefa{\v n}{\' a}k}},
  \bibinfo{author}{\bibfnamefont{V.}~\bibnamefont{Poto{\u c}ek}},
  \bibinfo{author}{\bibfnamefont{I.}~\bibnamefont{Jex}}, \bibnamefont{and}
  \bibinfo{author}{\bibfnamefont{C.}~\bibnamefont{Silberhorn}},
  \bibinfo{journal}{Science} \textbf{\bibinfo{volume}{336}},
  \bibinfo{pages}{55} (\bibinfo{year}{2012}).

\bibitem[{\citenamefont{Pittman and Franson}(2002)}]{bib:Pittman02}
\bibinfo{author}{\bibfnamefont{T.~B.} \bibnamefont{Pittman}} \bibnamefont{and}
  \bibinfo{author}{\bibfnamefont{J.~D.} \bibnamefont{Franson}},
  \bibinfo{journal}{Phys. Rev. A} \textbf{\bibinfo{volume}{66}},
  \bibinfo{pages}{062302} (\bibinfo{year}{2002}).

\bibitem[{\citenamefont{Dovrat et~al.}(2012)\citenamefont{Dovrat, Bakstein,
  Istrati, Shaham, and Eisenberg}}]{12dovrat2266}
\bibinfo{author}{\bibfnamefont{L.}~\bibnamefont{Dovrat}},
  \bibinfo{author}{\bibfnamefont{M.}~\bibnamefont{Bakstein}},
  \bibinfo{author}{\bibfnamefont{D.}~\bibnamefont{Istrati}},
  \bibinfo{author}{\bibfnamefont{A.}~\bibnamefont{Shaham}}, \bibnamefont{and}
  \bibinfo{author}{\bibfnamefont{H.~S.} \bibnamefont{Eisenberg}},
  \bibinfo{journal}{Optics Express} \textbf{\bibinfo{volume}{20}},
  \bibinfo{pages}{2266} (\bibinfo{year}{2012}).

\bibitem[{\citenamefont{Mauerer et~al.}(2009)\citenamefont{Mauerer, Avenhaus,
  Helwig, and Silberhorn}}]{09mauerer053815}
\bibinfo{author}{\bibfnamefont{W.}~\bibnamefont{Mauerer}},
  \bibinfo{author}{\bibfnamefont{M.}~\bibnamefont{Avenhaus}},
  \bibinfo{author}{\bibfnamefont{W.}~\bibnamefont{Helwig}}, \bibnamefont{and}
  \bibinfo{author}{\bibfnamefont{C.}~\bibnamefont{Silberhorn}},
  \bibinfo{journal}{Phys. Rev. A} \textbf{\bibinfo{volume}{80}},
  \bibinfo{pages}{053815} (\bibinfo{year}{2009}).

\bibitem[{\citenamefont{Barbieri et~al.}(2009)\citenamefont{Barbieri, Weinhold,
  Lanyon, Gilchrist, Resch, Almeida, and White}}]{09barbieri209}
\bibinfo{author}{\bibfnamefont{M.}~\bibnamefont{Barbieri}},
  \bibinfo{author}{\bibfnamefont{T.~J.} \bibnamefont{Weinhold}},
  \bibinfo{author}{\bibfnamefont{B.~P.} \bibnamefont{Lanyon}},
  \bibinfo{author}{\bibfnamefont{A.}~\bibnamefont{Gilchrist}},
  \bibinfo{author}{\bibfnamefont{K.~J.} \bibnamefont{Resch}},
  \bibinfo{author}{\bibfnamefont{M.~P.} \bibnamefont{Almeida}},
  \bibnamefont{and} \bibinfo{author}{\bibfnamefont{A.~G.} \bibnamefont{White}},
  \bibinfo{journal}{J. Mod. Opt.} \textbf{\bibinfo{volume}{56}},
  \bibinfo{pages}{209} (\bibinfo{year}{2009}).

\bibitem[{\citenamefont{Takesue et~al.}(2007)\citenamefont{Takesue, Tokura,
  Fukuda, Tsuchizawa, Watanabe, Yamada, and Itabashi}}]{Takesue2007}
\bibinfo{author}{\bibfnamefont{H.}~\bibnamefont{Takesue}},
  \bibinfo{author}{\bibfnamefont{Y.}~\bibnamefont{Tokura}},
  \bibinfo{author}{\bibfnamefont{H.}~\bibnamefont{Fukuda}},
  \bibinfo{author}{\bibfnamefont{T.}~\bibnamefont{Tsuchizawa}},
  \bibinfo{author}{\bibfnamefont{T.}~\bibnamefont{Watanabe}},
  \bibinfo{author}{\bibfnamefont{K.}~\bibnamefont{Yamada}}, \bibnamefont{and}
  \bibinfo{author}{\bibfnamefont{S.}~\bibnamefont{Itabashi}},
  \bibinfo{journal}{App. Phys. Lett.} \textbf{\bibinfo{volume}{91}},
  \bibinfo{pages}{201108} (\bibinfo{year}{2007}).

\bibitem[{\citenamefont{Silverstone et~al.}(2014)\citenamefont{Silverstone,
  Santagati, and Strain}}]{Silverstone2014}
\bibinfo{author}{\bibfnamefont{J.}~\bibnamefont{Silverstone}},
  \bibinfo{author}{\bibfnamefont{R.}~\bibnamefont{Santagati}},
  \bibnamefont{and} \bibinfo{author}{\bibfnamefont{M.}~\bibnamefont{Strain}}
  (\bibinfo{year}{2014}), \eprint{arXiv:1410.8332v4}.

\bibitem[{\citenamefont{Marsili et~al.}(2013)\citenamefont{Marsili, Verma,
  Stern, and Harrington}}]{marsili2013}
\bibinfo{author}{\bibfnamefont{F.}~\bibnamefont{Marsili}},
  \bibinfo{author}{\bibfnamefont{V.}~\bibnamefont{Verma}},
  \bibinfo{author}{\bibfnamefont{J.}~\bibnamefont{Stern}}, \bibnamefont{and}
  \bibinfo{author}{\bibfnamefont{S.}~\bibnamefont{Harrington}},
  \bibinfo{journal}{Nature Photonics} \textbf{\bibinfo{volume}{7}},
  \bibinfo{pages}{210} (\bibinfo{year}{2013}).

\bibitem[{\citenamefont{Yamashita et~al.}(2013)\citenamefont{Yamashita, Miki,
  Terai, and Wang}}]{Yamashita:13}
\bibinfo{author}{\bibfnamefont{T.}~\bibnamefont{Yamashita}},
  \bibinfo{author}{\bibfnamefont{S.}~\bibnamefont{Miki}},
  \bibinfo{author}{\bibfnamefont{H.}~\bibnamefont{Terai}}, \bibnamefont{and}
  \bibinfo{author}{\bibfnamefont{Z.}~\bibnamefont{Wang}},
  \bibinfo{journal}{Opt. Express} \textbf{\bibinfo{volume}{21}},
  \bibinfo{pages}{27177} (\bibinfo{year}{2013}).

\bibitem[{\citenamefont{Morris et~al.}(2014)\citenamefont{Morris,
  Francis-Jones, Wilcox, Tropper, and Mosley}}]{Morris2014}
\bibinfo{author}{\bibfnamefont{O.~J.} \bibnamefont{Morris}},
  \bibinfo{author}{\bibfnamefont{R.~J.} \bibnamefont{Francis-Jones}},
  \bibinfo{author}{\bibfnamefont{K.~G.} \bibnamefont{Wilcox}},
  \bibinfo{author}{\bibfnamefont{A.~C.} \bibnamefont{Tropper}},
  \bibnamefont{and} \bibinfo{author}{\bibfnamefont{P.~J.}
  \bibnamefont{Mosley}}, \bibinfo{journal}{Opt. Comm.}
  \textbf{\bibinfo{volume}{327}}, \bibinfo{pages}{39} (\bibinfo{year}{2014}).

\bibitem[{\citenamefont{Jin et~al.}(2014)\citenamefont{Jin, Shimizu, Morohashi,
  Wakui, Takeoka, Izumi, Sakamoto, Fujiwara, Yamashita, Miki et~al.}}]{Jin2014}
\bibinfo{author}{\bibfnamefont{R.-B.} \bibnamefont{Jin}},
  \bibinfo{author}{\bibfnamefont{R.}~\bibnamefont{Shimizu}},
  \bibinfo{author}{\bibfnamefont{I.}~\bibnamefont{Morohashi}},
  \bibinfo{author}{\bibfnamefont{K.}~\bibnamefont{Wakui}},
  \bibinfo{author}{\bibfnamefont{M.}~\bibnamefont{Takeoka}},
  \bibinfo{author}{\bibfnamefont{S.}~\bibnamefont{Izumi}},
  \bibinfo{author}{\bibfnamefont{T.}~\bibnamefont{Sakamoto}},
  \bibinfo{author}{\bibfnamefont{M.}~\bibnamefont{Fujiwara}},
  \bibinfo{author}{\bibfnamefont{T.}~\bibnamefont{Yamashita}},
  \bibinfo{author}{\bibfnamefont{S.}~\bibnamefont{Miki}}, \bibnamefont{et~al.},
  \bibinfo{journal}{Sci. Rep.} \textbf{\bibinfo{volume}{4}},
  \bibinfo{pages}{7468} (\bibinfo{year}{2014}).

\bibitem[{\citenamefont{Agrawal}(2012)}]{AgrawalNFO}
\bibinfo{author}{\bibfnamefont{G.~P.} \bibnamefont{Agrawal}},
  \emph{\bibinfo{title}{Nonlinear Fiber Optics, 5th edition}}
  (\bibinfo{publisher}{Academic Press Boston}, \bibinfo{year}{2012}).

\bibitem[{\citenamefont{Calkins et~al.}(2013)\citenamefont{Calkins, Mennea,
  Lita, Metcalf, Kolthammer, Lamas-Linares, Spring, Humphreys, Mirin, Gates
  et~al.}}]{Calkins:13}
\bibinfo{author}{\bibfnamefont{B.}~\bibnamefont{Calkins}},
  \bibinfo{author}{\bibfnamefont{P.~L.} \bibnamefont{Mennea}},
  \bibinfo{author}{\bibfnamefont{A.~E.} \bibnamefont{Lita}},
  \bibinfo{author}{\bibfnamefont{B.~J.} \bibnamefont{Metcalf}},
  \bibinfo{author}{\bibfnamefont{W.~S.} \bibnamefont{Kolthammer}},
  \bibinfo{author}{\bibfnamefont{A.}~\bibnamefont{Lamas-Linares}},
  \bibinfo{author}{\bibfnamefont{J.~B.} \bibnamefont{Spring}},
  \bibinfo{author}{\bibfnamefont{P.~C.} \bibnamefont{Humphreys}},
  \bibinfo{author}{\bibfnamefont{R.~P.} \bibnamefont{Mirin}},
  \bibinfo{author}{\bibfnamefont{J.~C.} \bibnamefont{Gates}},
  \bibnamefont{et~al.}, \bibinfo{journal}{Opt. Express}
  \textbf{\bibinfo{volume}{21}}, \bibinfo{pages}{22657} (\bibinfo{year}{2013}).

\end{thebibliography}

\end{document}